\documentclass[11pt,a4paper]{article}
\pdfoutput = 1
\usepackage{jheppub}
\usepackage{enumitem}
\usepackage{amsmath}
\usepackage{amsfonts}
\usepackage{graphicx}

\title{Near-Hagedorn Thermodynamics and Random Walks: a General Formalism in Curved Backgrounds}

\author[a]{Thomas G. Mertens,}
\author[a]{Henri Verschelde}
\author[b]{and Valentin I. Zakharov}
\affiliation[a]{Ghent University, Department of Physics and Astronomy\\
Krijgslaan, 281-S9, 9000 Gent, Belgium}
\affiliation[b]{ITEP, B. Cheremushkinskaya 25, Moscow, 117218 Russia,\\
Max-Planck Institut f¨ur Physik, 80805 M¨unchen, Germany,\\
Moscow Inst Phys \& Technol, Dolgoprudny, Moscow Region, 141700 Russia 
}
\emailAdd{thomas.mertens@ugent.be}
\emailAdd{henri.verschelde@ugent.be}
\emailAdd{vzakharov@itep.ru}
\abstract{In this paper we discuss near-Hagedorn string thermodynamics starting from the explicit path integral derivation found by \cite{Kruczenski:2005pj}. Their result is extended and the validity is checked by comparing with some known exact results. We compare this approach with the first-quantized one-loop result from the field theory action and establish correction terms to the above result.
}

\begin{document}

\maketitle

\section{Introduction}

It is an old fact that the behavior of string theory at high temperatures is rather different than that of a conventional thermodynamic system \cite{Atick:1988si}\cite{Horowitz:1997jc}\cite{Barbon:2004dd}. \\
Consider a gas of strings in a box with a fixed total amount of energy (microcanonical ensemble) \cite{Mitchell:1987hr}\cite{Mitchell:1987th}\cite{Bowick:1989us}\cite{Deo:1989bv}. As one increases the energy gradually, nothing special happens until suddenly most of the strings in the gas coalesce and form highly excited long strings. All energy pumped into the system goes into the excitation modes of the long string(s) and not in increasing the temperature. Actually the story is a bit more involved: a single string at high energy behaves as a long random walk. Whether the entire string gas is dominated by one or many long strings depends on the number of non-compact dimensions as is extensively discussed in \cite{Deo:1989bv}.\footnote{For 1 or 2 spatial non-compact dimensions, the gas is not dominated by long strings at all.} One can quickly see that long strings are favored for entropic reasons by using simple random walk arguments (see e.g. \cite{Zwiebach:2004tj}).\footnote{See also \cite{Manes:2004nd} for scattering amplitude arguments in favor of a random walk interpretation of the highly excited string.} Important to note is that the single string density of states always shows long random walk behavior and it is this aspect that we will consider. Whether the gas then is dominated by one string is another problem we shall not consider here.\\
Now consider the same story, but from a canonical ensemble point of view. Increasing the temperature to a critical value, causes the partition function to diverge due to the high density of highly excited string modes. This ultimate temperature is the so-called \emph{Hagedorn temperature} \cite{Hagedorn} and the canonical ensemble is not useable at higher temperatures. \\
A few years ago, the authors of \cite{Kruczenski:2005pj} made an explicit derivation of the link between these two approaches by deriving the single string random walk picture directly from the canonical ensemble using the string path integral at genus one. \\
This phenomenon also has a third different manifestation. Thermodynamics on any spacetime can be calculated on the so-called `thermal manifold' by Wick rotating the time coordinate and periodically identifying this coordinate. For particles nothing dramatic happens when doing this, strings however can wrap this Euclideanized time direction. The divergence manifests itself here by the masslessness of a winding string state at the critical temperature (and it becomes tachyonic when further heating the system). This string field is what is known as the \emph{thermal scalar} and when nearing the Hagedorn temperature, this field dominates the thermodynamics in the same way that the single string dominates the microcanonical picture \cite{Atick:1988si}. This field effectively represents the large density of states of highly excited string states. One should remark that this field is not a real field corresponding to physical particles but is an effective field theory degree of freedom which dominates the string thermodynamics at high energy.

This random walk picture also arises in black hole geometries \cite{Susskind:1993ws}\cite{Susskind:2005js}. The long string surrounds the event horizon and forms the stretched horizon (or the black hole membrane as it is called in the earlier literature\footnote{See e.g. \cite{Thorne:1986iy} and references therein.}). This work started with the question: `Can we apply the methods developed in \cite{Kruczenski:2005pj} to the black hole case?'
In the case of a black hole, the local temperature increases as one approaches the horizon and at a distance of the order of the string length $\sqrt{\alpha'}$, it exceeds the flat space Hagedorn temperature. One expects a condensate of winding tachyons close to the horizon which in the Lorentzian case is responsible for the appearance of a stretched horizon. Strong evidence for this scenario was shown using the exactly solvable 2D black hole or cigar \cite{Kutasov:2005rr} appearing in Little String Theory \cite{Kutasov:2000jp}.
A relation between condensed winding modes and black hole entropy has also been discussed in \cite{Dabholkar:2001if} using $\mathbb{C}/\mathbb{Z}_n$ orbifolds. The idea \cite{Adams:2001sv} is that the tachyons condense at the tip of the cone and relax the cone to flat space. In this case, closed string field theory can be used because of the localized nature of the winding tachyons \cite{Okawa:2004rh}. For Schwarzschild black holes, the situation is less clear \cite{Kutasov:2005rr}.
Recently, research on the nature of the stretched horizon has been rekindled in \cite{Almheiri:2012rt} who propose the existence of a firewall.\footnote{See \cite{Braunstein:2009my} for an earlier development in this direction.} As pointed out in \cite{Giveon:2012kp}\cite{Giveon:2013ica}, the existence of a winding string zero mode close to the horizon is a possible stringy realization of this idea and deserves further study. Our endeavor is therefore to develop from the string path integral, a general framework for the thermal scalar in curved spacetime backgrounds.

Before we arrive there however, we will first reanalyze the derivation of \cite{Kruczenski:2005pj} and extend and test their result in several situations that are easier to understand than black hole horizons. We discuss the application to black hole horizons themselves in \cite{Mertens:2013zya}. Several other examples will be postponed to a companion paper \cite{examples}. 
We want to analyze the random walk picture of highly excited strings in general backgrounds from the canonical ensemble and see if we can get aspects of the above picture out of it. \\

This paper is organized as follows.\\
In section \ref{pathderiv} we review and extend the path integral derivation of the random walk behavior in general backgrounds as was put forward by \cite{Kruczenski:2005pj}. The beauty of this path integral approach is the physical picture of a random walk that clearly emerges once the dust settles. We comment on several of the difficulties that appear in our derivation. The most important of these is that we seem to miss several terms in the resulting particle action, indicating that we did not take the near-Hagedorn limit correctly.\\
In section \ref{comp} we compare the results from the second section to some explicitly known flat spacetime results. We will see in these explicit examples that we do indeed reproduce the expected results if we include a correction corresponding to the flat space tachyon mass in the action. \\
In section \ref{corrections} we will calculate the correction terms explicitly in flat spacetime while taking a path integral perspective (i.e. without comparing to known results). This will demonstrate where the above term comes from. \\
We take a different point of view in section \ref{alternative} and try to see whether we can make contact with one-loop results of field theory actions. The reason we take this approach is because in this case it is computationally easier to deal with field theory actions than to manipulate path integral expressions. We will find a match for flat backgrounds but for general backgrounds we find other terms as well in the action. These are terms arising from the $\sqrt{G_{00}}$ metric component in the measure in the field theory action. We interpret these as other terms we missed in the derivation of the second section. \\
Several technical computations are given in the appendices.
Examples of these methods are presented in companion papers \cite{Mertens:2013zya}\cite{examples}.

\section{Path integral approximation for dominance by singly wound strings}
\label{pathderiv}
The authors of \cite{Kruczenski:2005pj} have given an explicit path integral picture of the thermal scalar. In this section we review and extend their derivation of the random walk picture of highly excited strings (while making some modifications near the end). \\
The goal is to derive the free energy of a gas of non-interacting strings in a curved background in the limit where highly excited strings dominate (in the microcanonical ensemble). In the canonical ensemble this corresponds to temperatures near the Hagedorn temperature of the specific background. Let us first remark that the relation between the microcanonical and canonical ensemble is not entirely clear in string theory: several conceptual problems arise due to the asymptotic exponential degeneracy of states \cite{Mitchell:1987th}. In what follows we will perform our computations in the canonical ensemble and leave further study of this issue in our case to future work. A partial motivation for this is that the canonical ensemble is often used as a starting point to compute the relevant microcanonical quantities \cite{Brandenberger:1988aj}\cite{Deo:1988jj}.

\subsection{The thermal manifold to calculate string thermodynamics}

Thermodynamics in a general background depends obviously on the choice of time variable. So to describe e.g. the free energy we first have to choose a preferred time coordinate and then calculate the free energy associated to that specific time coordinate.\\
The starting point of the derivation is the \emph{assumption} that the free energy of a non-self-interacting string gas in a certain time-independent background (defined as the sum of the free energies of the individual particle states in the string spectrum) is proportional to the torus partition function of a single string on the thermal manifold (same background, but with $X^0$ Wick-rotated and compactified with length $\beta$, the inverse temperature). So
\begin{equation}
\label{cruc}
Z_{1 string} = -\beta F_{string gas}.
\end{equation}
We say non-self-interacting since the string gas does interact with the background, but it does not interact with itself (reflected in the fact that we only consider the torus amplitude). \\
We make the following comments regarding this assumption:
\begin{itemize}
\item{It holds for the flat bosonic case as proven by Polchinski \cite{Polchinski:1985zf}.}
\item{It holds also for toroidal compactifications of the flat bosonic string. As an example, we prove such a statement in \cite{examples}.}
\item{In \cite{Maldacena:2000kv} this equality was used in $AdS_3$ to identify the string spectrum with the proposed spectrum obtained from harmonic analysis \cite{Maldacena:2000hw}.}
\item{The authors of \cite{Ferrer:1990na} prove that the analogous statement holds for open strings on the cylinder worldsheet in a constant background electromagnetic field.}
\item{For flat space superstrings and heterotic strings, such a statement also holds \cite{Alvarez:1986sj} but one needs to be careful in the interplay between the bosonic boundary conditions and the fermionic boundary conditions (spin structure), i.e. the GSO projection.}
\end{itemize}

\subsection{Deriving the thermal scalar}
\label{deriv}

We will now explore the thermodynamics of closed strings at the one loop level (genus 1). \\

We must first be a bit more precise on the relation (\ref{cruc}). The modular integration of the torus amplitude is chosen to be the entire strip \cite{Polchinski:1985zf} and we restrict the Euclidean time coordinate to winding around only one torus cycle. In a second stage, in flat space, one can use the theorem by \cite{McClain:1986id}\cite{O'Brien:1987pn} to relate this to a modular integral over the fundamental domain, while replacing the zero-mode sum over a single quantum number by a double sum over both momenta and winding. For superstrings, an extension of this theorem needs to be used. The logic is basically the same: if we start with the fundamental domain, we restrict the double sum to a single sum and extend the modular domain to the entire strip (see e.g. \cite{Kutasov:2000jp} for an example of such a procedure). 
In what follows we view the strip domain as the relevant one for thermodynamics and restrict the Euclidean time coordinate to a single torus cycle. For now, we also assume that no other coordinates are compactified. We will extend this assumption in \cite{examples}.\\

We are interested in the dominant contribution and so we restrict ourselves to winding $\pm 1$ around the Euclidean time direction. At least in spaces where the thermal circle is topologically stable, we expect strings that are wrapped multiple times to be more massive. Indeed, in the flat space string spectrum, the winding $\pm 1$ mode becomes massless at the Hagedorn temperature \cite{Atick:1988si} and strings with higher winding numbers are massive. The zero-winding modes correspond to the zero-temperature vacuum energy and we are not interested in this here. \\
This intuition is well-founded for spaces with topologically stable thermal circles, but what about other spaces? We know Euclidean black hole backgrounds are cigar-shaped and the thermal circle shrinks to zero size at the horizon. Is there still a dominating winding mode present? 
For now we will \emph{assume} that indeed winding $\pm 1$ modes are dominant and we focus on them. We present examples of these phenomena in a companion paper \cite{examples}.\\

\begin{figure}[h]
\centerline{\includegraphics[width=0.8\textwidth]{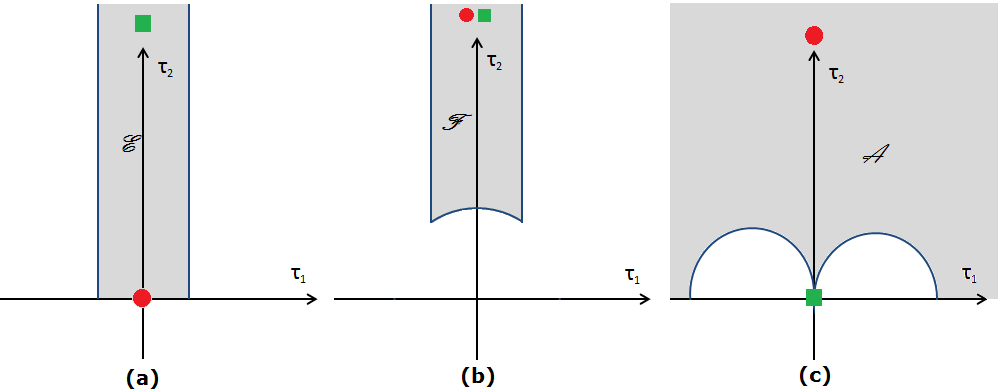}}
\caption{A circle represents the winding tachyon divergence. A square represents the closed string tachyon divergence (or the massless states for superstrings). (a) Free energy evaluated in the strip. (b) Free energy as a partition function of a single string on the thermal manifold. The equivalence with (a) follows from the theorem in \cite{McClain:1986id}\cite{O'Brien:1987pn}. (c) Free energy after the modular transformation $\tau \to -\frac{1}{\tau}$ on (a).}
\label{fund}
\end{figure}

We start from the following torus path integral in an external field $G_{\mu\nu}$ in $D$ spacetime dimensions: 
\begin{equation} 
Z_{T_2} = \int_0^\infty \frac{d\tau_2}{2\tau_2} \int_{-1/2}^{1/2} d\tau_1 \Delta_{FP} \int \left[\mathcal{D}X\right]\sqrt{G}
\exp -\frac{1}{4\pi\alpha'} \int d^2\sigma \sqrt h h^{\alpha \beta} \partial_\alpha X^\mu \partial_\beta X^\nu G_{\mu\nu}(X).
\end{equation}
where $\Delta_{FP}$ denotes the Faddeev-Popov determinant from the (Diff $\times$ Weyl) gauge-fixing procedure. We choose Euclidean signature for both the worldsheet and the target space manifold. The worldsheet metric has been fixed to
\begin{equation}
h_{\alpha\beta} = \left[\begin{array}{cc} 
1 & \tau_1  \\
\tau_1 & \tau_1^2+\tau_2^2 \end{array}\right],
\end{equation}
and the torus worldsheet is represented by a square with sides equal to 1.
We consider strings that are singly wound around the Euclidean time direction:
\begin{align}
X^\mu(\sigma_1+1,\sigma_2) & =  X^\mu(\sigma_1,\sigma_2), \quad \mu = 0\hdots D-1,\nonumber \\
X^i(\sigma_1,\sigma_2+1) & =  X^i(\sigma_1,\sigma_2), \quad i=1 \hdots D-1,\nonumber \\
X^0(\sigma_1,\sigma_2+1) & =  X^0(\sigma_1,\sigma_2) \pm \beta.
\end{align}
where the wrapping is around the temporal worldsheet coordinate.
The Hagedorn divergence is due to the $\tau_2 \rightarrow 0$ (ultraviolet) behavior of the torus path integral (as shown in figure \ref{fund}(a)). 
The essential idea of the thermal scalar is that this ultraviolet divergence for $\tau_2 \rightarrow 0$ can be described through a UV/IR connection as an infrared divergence for $\tau_2 \rightarrow \infty$. 
In standard approaches \cite{Atick:1988si}, one uses the link between the strip and the fundamental domain to make this correspondence. The divergence appears there as a string state that becomes massless at precisely the Hagedorn temperature (see figure \ref{fund}(b)). However, we follow \cite{Kruczenski:2005pj} and instead use a modular transformation on the strip domain (figure \ref{fund}(c)). This will enable us to deal with the integral over $\tau_1$ later on.
So, qualitatively, the ultraviolet divergence becomes an infrared divergence (due to the massless thermal scalar). More precisely, we perform the transformation $\tau \to -\frac{1}{\tau}$ in the modular integral and then swap the roles of $\sigma_1$ and $\sigma_2$. As a consequence, the string wraps the thermal circle along the spatial worldsheet coordinate $\sigma_1$: 
\begin{align}
X^\mu(\sigma_1,\sigma_2+1) & = X^\mu(\sigma_1,\sigma_2), \quad \mu = 0\hdots D-1,\nonumber\\
X^i(\sigma_1+1,\sigma_2) & = X^i(\sigma_1,\sigma_2), \quad i =1\hdots D-1,\nonumber\\
X^0(\sigma_1+1,\sigma_2) & = X^0(\sigma_1,\sigma_2) \pm \beta.
\end{align}

In figure \ref{fund}(a), there is a UV divergence caused by the exponentially growing string density. This corresponds to long string dominance. For bosonic strings, after using the theorem from \cite{McClain:1986id}\cite{O'Brien:1987pn}, we get the picture of figure \ref{fund}(b). Here we interpret this divergence as the tachyonic character of the singly wound string. To extract solely the winding tachyon contribution, we use a modular transformation in the strip (a). This displaces the winding tachyon divergence to $\tau_2 \to \infty$ (figure \ref{fund}(c)).\footnote{Strictly speaking, this is only valid for $\tau_1=0$. Nevertheless, we will obtain the expected results.} \\

Let us now consider this dual (in the modular sense $\tau \rightarrow -\frac{1}{\tau}$) path integral for $\tau_2 \rightarrow \infty$. 
We can use the reparametrization invariance of the path integral to define new worldsheet coordinates $(\sigma,\tau)$:
\begin{equation}
\left\{ \begin{array}{l} \sigma= \frac{\sigma_1}{\tau_2}, \tau = \sigma_2 \\ X(\sigma_1,\sigma_2) \rightarrow X(\sigma,\tau). \end{array} \right.
\end{equation}
The worldsheet action now becomes using the torus metric: 
\begin{eqnarray}
\label{action1}
S = \frac{1}{4\pi \alpha'} \left[ \left(1 + \frac{\tau_1^2}{\tau_2^2}\right) \int_0^{1/\tau_2} d\sigma \int_0^1 d\tau G_{\mu\nu} \partial_\sigma X^\mu \partial_\sigma X^\nu \right. \nonumber \\
\left.
+ 2 \frac{\tau_1}{\tau_2} \int_0^{1/\tau_2} d\sigma \int_0^1 d\tau G_{\mu\nu} \partial_\sigma X^\mu \partial_\tau X^\nu 
+ \int_0^{1/\tau_2} d\sigma \int_0^1 d\tau G_{\mu\nu} \partial_\tau X^\mu \partial_\tau X^\nu \right].
\end{eqnarray} 
We next consider a Fourier series expansion in the $\sigma$ worldsheet coordinate. 
\begin{align}
X^i(\sigma,\tau) & = \sum_{n=-\infty}^{+\infty} e^{i(2\pi n \tau_2) \sigma} X_n^i(\tau), \\
X^0(\sigma,\tau) & = \pm \beta \tau_2 \sigma +  \sum_{n=-\infty}^{+\infty} e^{i(2\pi n \tau_2) \sigma} X_n^0(\tau).
\end{align}
In the $\tau_2 \rightarrow \infty$ limit, only the $n=0$ mode survives ($\langle (X_n^\mu)^2\rangle \sim 1/\tau_2$ for $n\neq 0$) and we get a dimensional reduction from a two dimensional non-linear $\sigma$-model on the worldsheet to quantum mechanics on the worldline and the string theory reduces to a particle theory (of the thermal scalar). Physically this corresponds to neglecting the temporal worldsheet dependence of the string which we started with.\\
Defining 
\begin{equation}
X_0^i(\tau) = X^i (\tau), \quad X_0^0(\tau) = X^0 (\tau),
\end{equation}
the particle action becomes: 
\begin{eqnarray} 
S_{part} =  \frac{1}{4\pi \alpha' \tau_2} \left[ \beta^2(\tau_1^2 + \tau_2^2)  \int_0^1 d\tau G_{00} \pm 2 \tau_1 \beta \int_0^1 d\tau G_{00} \partial_\tau X^0 \right. \nonumber \\
\qquad \qquad\qquad\qquad
\left.\pm 2 \tau_1 \beta \int_0^1 d\tau G_{0i} \partial_\tau X^i +
\int_0^1 d\tau G_{\mu\nu} \partial_\tau X^\mu \partial_\tau X^\nu \right]
\end{eqnarray}
for winding number $\pm1$. 

Of course, in dimensional reduction, one always loses information about the high energy degrees of freedom on the worldsheet, which could be important. The situation is similar to dimensional reduction in high temperature field theory where one loses the perturbative Stefan-Boltzmann result coming from the high energy degrees of freedom \cite{Braaten:1995jr}\cite{Braaten:1995cm}. In a sense, one can view $\tau_2$ as a `spatial' worldsheet temperature.
The dimensional reduction has replaced a string path integral with a particle path integral. 
So the lost degrees of freedom are the orthogonal oscillations of the string. \\

In the case of bosonic strings in flat space ($G_{\mu\nu} = \delta_{\mu\nu}$), the orthogonal oscillations (including the Faddeev-Popov factor $\Delta_F \tau_2^{-1}$) give a factor: 
\begin{equation}
 |\eta(\tau)|^{-48} \exp -\frac{\pi R^2}{\tau_2 \alpha'} |n\tau - m|^2 ,
\end{equation}
for one compact dimension with radius $R$. The symbol $\eta$ denotes the Dedekind $\eta$-function.
Using the modular transformation and taking the limit $\tau_2 \to \infty$ yields
\begin{equation}
e^{4\pi \tau_2} e^{-\frac{\beta^2}{4\pi \alpha' \tau_2} (\tau_1^2+\tau_2^2)}.
\end{equation}
Comparing with $S_{part}$, we find that we have to add 
\begin{equation}
\label{bosAdd}
\Delta S = -4\pi \tau_2
= -\frac{\tau_2^2 \beta_H^2}{4\pi \alpha' \tau_2}.
\end{equation}
This correction term was only calculated for flat spacetime. We expect other corrections when we consider a generally curved background. One of the goals of this paper is precisely to get a handle on these corrections. We will have more to say about this further on. \\

Introducing the parameter $t=\tau_2 \tau$ and adding $\Delta S$, the particle action is finally 
\begin{equation}
\label{particleaction}
S_{part} =  \frac{1}{4\pi \alpha' } \left[ \beta^2\frac{\left|\tau\right|^2}{\tau_2^2} \int_0^{\tau_2} dt G_{00} - \beta_H^2 \tau_2 \pm 2 \frac{\tau_1}{\tau_2} \beta \int_0^{\tau_2} dt G_{0\mu} \partial_t X^\mu 
+\int_0^{\tau_2} dt G_{\mu\nu} \partial_t X^\mu \partial_t X^\nu \right].
\end{equation}
If the metric is time independent, whatever the corrections are, the $X^0$ integration is Gaussian in this case and can be integrated out exactly. 
From now on we set $G_{0i}=0$. For the Gaussian $X^0$ integration ($G_{00}$ is independent of $X^0$) we first solve the classical equation: 
\begin{equation}
\partial_t \left[ \left( G_{00}(\vec X) \partial_t X^{0,cl}(t) \right) \pm \frac{\tau_1}{\tau_2} \beta G_{00}(\vec X) \right] = 0
\end{equation} 
or 
\begin{equation} 
G_{00}(\vec X) \partial_t X^{0,cl} \pm \frac{\tau_1}{\tau_2} \beta G_{00}(\vec X) = C.
\end{equation} 
The constant $C$ is determined by periodicity: 
\begin{equation}
\int_0^{\tau_2} \partial_t X^0 dt = 0 
\end{equation}
so that 
\begin{equation}
C = \pm \frac{ \tau_1 \beta}{\left \langle 1/G_{00} \right \rangle},
\end{equation}
where we denoted $\langle A \rangle = \int_{0}^{\tau_2}A dt.$
The classical action (of the $X^{0}$-dependent contributions) is
\begin{equation} 
S\left(X^{0,cl}\right) =  \frac{1}{4\pi \alpha' } \left[ \frac{ \tau_1^2 \beta^2}{\left \langle 1/G_{00} \right \rangle} - \frac{\tau_1^2}{\tau_2^2}\beta^2 \langle G_{00} \rangle \right].
\end{equation}
The second term in the classical action cancels the $\tau_1^2$ term in $S_{part}$.
Therefore, putting $X^0 = X^{cl} + \tilde X^0$, we find: 
\begin{equation} 
S_p =  \frac{1}{4\pi \alpha'} \left[ \frac{ \tau_1^2 \beta^2}{\left \langle 1/G_{00} \right \rangle} + \beta^2 \int_0^{\tau_2} dt G_{00} - \beta_H^2 \tau_2 +  \int_0^{\tau_2} dt G_{00} (\partial_t \tilde X^0)^2 + \int_0^{\tau_2} dt G_{ij} \partial_t X^i \partial_t X^j \right].
\end{equation}
The $\tilde X^0$ path integral is: 
\begin{equation} 
Z_0 = \int \left[ \mathcal{D}\tilde X^0 \right] \prod_t \sqrt{G_{00}(\vec X(t))} \exp - \frac{1}{4\pi \alpha'} \int_0^{\tau_2} dt G_{00}(\vec X) (\partial_t \tilde X^0)^2.
\end{equation}
Using the following identity for one timestep $\epsilon$:
\begin{align} 
\int_{-\infty}^{+\infty} du \sqrt{\frac{a_1}{\pi}} &\exp \left(- \frac{a_1}{\epsilon}(x-u)^2\right) \sqrt{\frac{a_2}{\pi}} \exp \left(- \frac{a_2}{\epsilon}(u-y)^2\right) \nonumber \\
&=
\sqrt \epsilon \sqrt{\left(\frac{a_1 a_2}{a_1+a_2}\right) \frac{1}{\pi}} \exp \left(- \frac{a_1 a_2}{a_1+a_2} \frac{(x-y)^2}{\epsilon}\right),
\end{align}
we find after $n$ intermediate timesteps: 
\begin{equation} 
\left(\sqrt \epsilon\right)^n \sqrt{\left(\frac{1}{\sum_{i=1}^n 1/a_i}\right) \frac{1}{\pi}} 
\exp \left(- \frac{1}{\sum_{i=1}^n 1/a_i} \frac{(x-y)^2}{\epsilon}\right).
\end{equation}
Using 
\begin{equation}
a_i =\frac{1}{4\pi \alpha'} G_{00}(t_i),
\end{equation}
and performing the final integration over $\tilde X_0(0) =\tilde X_0(\tau_2)$ which gives a factor $\beta$, we find 
\begin{equation} 
Z_0 = N \beta \sqrt{\frac{1}{\langle G_{00}^{-1}\rangle}},
\end{equation}
where $N$ is a normalization factor given by\footnote{By carefully taking the measure into account, we can determine the normalization factor $N$. We can determine this constant equally easy from the flat space limit (since the constant does not depend on the background) where the path integral is given by
\begin{equation}
Z_0 = \int \left[dX_0\right]\exp{-\frac{1}{4\pi\alpha'}}\int_{0}^{\tau_2}dt (\partial_t X_0)^{2} = \frac{\beta}{\sqrt{4\pi^2\alpha'\tau_2}}.
\end{equation}
Comparing the results immediately gives the above value for $N$.
}
\begin{equation}
N = \sqrt{\frac{1}{4\pi^2\alpha'}}.
\end{equation}
Note that the appearance of $\prod_t \sqrt{G_{00}}$ in the measure is essential for this simple result. \\

The $\tau_1$ integration can also be performed exactly for $\tau_2 \rightarrow \infty$ because the dual domain of integration for $\tau_1$ is $]-\infty,+\infty[$. 
This gives the factor 
\begin{equation} 
\frac{2\pi \sqrt{\alpha'} \sqrt{\langle G_{00}^{-1}\rangle}}{\beta}.
\end{equation}
Putting everything together, the $\langle G_{00}^{-1}\rangle$ factors cancel and we finally obtain the following particle path integral after integrating out $X_0$: 
\begin{equation} 
Z_p = 2\int_0^\infty \frac{d \tau_2}{2\tau_2} \int \left[ \mathcal{D}X \right] \sqrt{\prod_{t} \det G_{ij}} \exp - S_p(X) 
\end{equation}
where 
\begin{equation}
S_p = \frac{1}{4\pi \alpha'}\left[ \beta^2 \int_0^{\tau_2} dt G_{00} - \beta_H^2 \tau_2 
+\int_0^{\tau_2} dt G_{ij} \partial_t X^i \partial_t X^j\right].
\end{equation}
The prefactor $2$ is the result of summing over both windings $\pm 1$, since these give equal contributions.
Note that we do not find the path integral measure $\sqrt{G_{00}  \det G_{ij}}$ in contrast to the authors of \cite{Kruczenski:2005pj}. \\

In all, we have reduced the full string partition function to a partition function for a non-relativistic particle moving in the purely spatial curved background. Because of the swapping of worldsheet coordinates, the time evolution of the particle in its random walk is equal to the spatial form of the long highly excited string. The long string in real spacetime has a shape described by the above random walk.\\

Before moving on, let us briefly recapitulate several of the delicate points in the previous derivation:
\begin{itemize}
\item{We assumed that the free energy of a string gas can be described by a path integral on the thermal manifold over the modular strip. As we already discussed, this is a very plausible statement. But, as far as we know, it has not been proven in general.}

\item{This picture provides an explicit realization of the correspondence between the long highly excited string and the dominant behavior of the canonical (single-string) partition function. The Euclidean time coordinate has been integrated out to explicity show the random walk behavior in the spatial submanifold. If we are interested in the configurations of the entire string gas, we should look at the multi-string partition function which is given by
\begin{equation}
Z_{mult} = e^{-\beta F}
\end{equation}
and it is clear that this has contributions from multiple random walks. Whether this is dominated by single random walks then requires an analysis analogous to \cite{Deo:1989bv} that we shall not consider here.}

\item{Note that since the final result does not depend on $X^{0}$ anymore, the Wick rotation (naively) appears trivial in this case. To put it another way, we end up with a random walk in the spatial submanifold and it appears irrelevant whether we view this as the spatial submanifold of the Lorentzian or Euclidean background. There is however one important influence of this Wick rotation: if one chooses a black hole background with $G_{00} \to 0$ as $r \to r_{H}$ (for instance a Schwarzschild black hole in Schwarzschild coordinates), the radial coordinate only lives outside the event horizon. When viewing this random walk as a walk in Lorentzian signature spacetime, this means the random walk cannot intersect the horizon and go into the interior of the black hole. This makes sense, since we are discussing thermodynamics as observed by fiducial observers and spacetime effectively ends at the horizon for these observers (cfr. the membrane paradigm \cite{Thorne:1986iy}). The concrete application to random walks near black hole horizons is discussed in \cite{Mertens:2013zya}.}

\item{We stated that the $\tau_2 \to$ 0 limit corresponding to the UV divergence, is mapped to the $\tau_2 \to \infty$ limit as a IR divergence. This only holds for $\tau_1 = 0$. We did not consider the effect of other values of $\tau_1$ on the derivation. This is a subtle point, but we will find further on that nevertheless we reproduce expected results whenever we can compare with other results.}

\item{In the worldsheet dimensional reduction, we ignored all higher order terms (we set them to zero) because they were not dominant in the large $\tau_2$ limit. A good approach would be to integrate all of these out and get corrections for the lowest Fourier mode. This would then give us the correction term that we missed. We previously found this by comparing our result with the known exact result. In what follows we will compare the previous result with several exact results from type II and heterotic superstrings and we will find perfect agreement. We will also exactly integrate out all higher Fourier modes in flat spacetime and indeed see that these provide us with the missed correction term in the action.}

\item{There is a divergence for $\tau_2 \to 0$ in this result. This was not present at first and was introduced `by hand' when we chose the lower boundary for the $\tau_2$ integral. In principle, we should \emph{not} take $0$ as the lower limit, since we assumed that $\tau_2$ was large. We chose this to find agreement with the field theory picture (see further) that indeed has lower value $0$. This is actually the same reasoning as discussed in \cite{Polchinski:1998rq} (chapter 7): if we rewrite the path integral to explicitly display all string contributions and then drop all but one of these states, we would get the same result as a field theory first-quantized vacuum loop (provided the integral boundaries are kept the same).
This divergence is not a consequence of a tachyon, but instead follows from the fact that we only retain one string field. The divergence that sets in is the usual field theory UV divergence. By dropping all other string excitations, we have lost the finiteness of the amplitudes. We are however only interested here in the $\tau_2 \to +\infty$ limit so this will not bother us. Also note that in the bosonic string case, the full string amplitude has a tachyon in the small $\tau_2$ limit, but when doing the manipulations in the large $\tau_2$ limit, we lose this tachyon divergence but instead get the field theory divergence.}
\end{itemize}

\subsection{Some extensions}
\label{exten}
Let us now make two extensions to the method discussed above: a background Kalb-Ramond field and general stationary spacetimes (so $G_{0i} \neq 0$).
\subsubsection*{Adding a Kalb-Ramond background}
We extend the previous result to contain a background Kalb-Ramond field. The starting point is the following addition to the string action
\begin{equation}
S_{extra} = \frac{1}{4\pi\alpha'}\int d^{2}\sigma \sqrt{h}i \epsilon^{ab}B_{\mu\nu}(X)\partial_{a}X^{\mu}\partial_{b}X^{\nu}.
\end{equation}
Since $\sqrt{h}\epsilon^{12} = 1$, the modular transformation has no effect on this term except the swapping of the roles of $\sigma_1$ and $\sigma_2$. We obtain
\begin{equation}
S_{extra} = -\frac{1}{2\pi\alpha'}\int d\sigma_1 d\sigma_2 i B_{\mu\nu}(X)\partial_{1}X^{\mu}\partial_{2}X^{\nu}.
\end{equation}
Doing the substitution to $\sigma$ and $\tau$ as before yields
\begin{equation}
\label{Bfield}
S_{extra} = -\frac{1}{2\pi\alpha'}\int_{0}^{1/\tau_2} d\sigma \int_{0}^{1}d\tau i B_{\mu\nu}(X)\partial_{\sigma}X^{\mu}\partial_{\tau}X^{\nu}.
\end{equation}
The next step is the Fourier expansion. There is only one non-zero possibility here and it finally gives the following addition to the particle action
\begin{equation}
S_{extra} = \mp i\frac{\beta}{2\pi\alpha'}\int_{0}^{\tau_2}dt B_{0i}(X)\partial_{t}X^{i}.
\end{equation}
This can be interpreted as the minimal coupling of a non-relativistic particle (with a suitably normalized charge) to a vector potential $A_{i} = B_{0i}$. So the resulting particle that lives in one dimension less than the original string, is also minimally coupled to an electromagnetic field. \\
Note that such a term is no longer symmetric under the positive and negative windings (as is obvious from the orientation reversal symmetry breaking property of the Kalb-Ramond field). In the particle path integral, the interpretation is that the particles are oppositely charged under the electromagnetic field.

\subsubsection*{Non-static spacetimes}
\label{nonstat}
One can readily extend the previous calculation to the case where $G_{0i} \neq 0$. So we consider stationary but non-static spacetimes (e.g. a string-corrected version of the Kerr black hole). We present the derivation in appendix \ref{AppAA}. One arrives at
\begin{equation} 
Z_p = 2\int_0^\infty \frac{d \tau_2}{2\tau_2} \int \left[ d\vec X \right] \sqrt{\prod \det \left(G_{ij} - \frac{G_{0i}G_{0j}}{G_{00}}\right)} \exp - S_p(\vec X) 
\end{equation}
where 
\begin{equation}
S_p = \frac{1}{4\pi \alpha'}\left[ \beta^2 \int_0^{\tau_2} dt G_{00} - \beta_H^2 \tau_2 
+\int_0^{\tau_2} dt \left(G_{ij} - \frac{G_{0i}G_{0j}}{G_{00}}\right) \partial_t X^i \partial_t X^j\right].
\end{equation}
Thus the only modification is a change in the spatial metric $G_{ij} \to G_{ij} - \frac{G_{0i}G_{0j}}{G_{00}}$. We will find this action again in section \ref{dimred} using dimensional reduction and T-duality.

\section{Comparison with exact results}
\label{comp} 
In the previous section we saw that for bosonic strings we needed to include a correction term in the action proportional to the Hagedorn temperature. We now look into the other types of strings. Just like we did for the bosonic string in the previous section, our strategy will be to compare the path integral result with the known exact expression for the free energy. 

\subsection{Flat space superstring}
\label{supers}
For the superstring, one might be initially worried about the effects of the worldsheet fermions on the previous derivation and especially their interplay with the worldsheet bosons (the GSO projection). In \cite{Alvarez:1986sj} it is shown that in thermal path integrals with the strip as the modular integration domain, one must choose only one spin structure for the fermions. 
We thus expect that the fermionic contribution will only be a modification of the Hagedorn temperature (at least for flat space). We now show that this is indeed the case.
The free energy for superstrings is given by the following expression \cite{Alvarez:1986sj}\footnote{One readily checks that this is the same expression as that obtained in \cite{Atick:1988si} when one restricts the expression for the free energy from \cite{Atick:1988si} to a single sum and extends the domain of integration to the entire strip.}
\begin{equation}
F = -2 V_9\int_{0}^{+\infty}d\tau_2\int_{-1/2}^{1/2}\frac{d\tau_1}{\tau_2^{6}(2\pi^2\alpha')^5}\left[\vartheta_3\left(0,\frac{i\beta^2}{4\pi^2\alpha'\tau_2}\right)-\vartheta_4\left(0,\frac{i\beta^2}{4\pi^2\alpha'\tau_2}\right)\right]\left|\vartheta_4(0,2\tau)\right|^{-16}.
\end{equation}
In appendix \ref{appsuper} we prove that in the limit of large $\tau_2$ (after the modular transformation) this is equal to
\begin{equation}
\label{freeEnsuper}
F = -2 V_9\iint_{\mathcal{A}} \frac{d\tau_2d\tau_1}{2\tau_2}\frac{1}{(4\pi^2\alpha'\tau_2)^5}e^{-\frac{\beta^2}{4\pi\alpha'}\frac{\tau_1^2+\tau_2^2}{\tau_2}}e^{2\pi\tau_2}.
\end{equation}
where the integration region $\mathcal{A}$ was shown in figure \ref{fund}(c).\\
The path integral from the previous section (\emph{without} a correction) is given by 
\begin{equation}
Z = 2\iint_{\mathcal{A}}\frac{d\tau_1d\tau_2}{2\tau_2}\int\left[\mathcal{D}X\right]e^{-\frac{\tau_1^2\beta^2}{4\pi\alpha'\tau_2}}e^{-\frac{\beta^2\tau_2}{4\pi\alpha'}}e^{-\frac{1}{4\pi\alpha'}\int_{0}^{\tau_2}dt\left(\partial_tX^\mu\right)^2}.
\end{equation}
Note that each flat space integral is given by
\begin{equation}
\int \left[\mathcal{D}X\right]e^{-\frac{1}{4\pi\alpha'}\int_{0}^{\tau_2}dt\left(\partial_tX^\mu\right)^2} = \frac{L}{\sqrt{4\pi^2\alpha'\tau_2}},
\end{equation}
where $L$ is the length of the space. So the path integration over the free coordinate fields reproduces precisely the $\frac{\beta V_9}{(4\pi^2\alpha'\tau_2)^5}$ factor and the factor $2$ in the beginning is the result of the sum over both windings. Considering finally the relation $Z = -\beta F$, we conclude that this is exactly the same as equation (\ref{freeEnsuper}) if we include a factor $e^{2\pi\tau_2}$. This corresponds to the superstring Hagedorn temperature $\beta_{H} = \pi \sqrt{8\alpha'}$. \\

So we learn that also the flat spacetime superstring gets the same type of correction as the flat spacetime bosonic string.

\subsection{Flat space heterotic string}
\label{heterotic}
We now look into heterotic strings in flat spacetime. For concreteness we focus on the $E_8 \times E_8$ string but the result is more general (see remarks further). The general expression for the free energy of the heterotic string is given by \cite{Alvarez:1986sj}
\begin{equation}
F = -2 V_9\int_{0}^{+\infty}d\tau_2\int_{-1/2}^{1/2}\frac{d\tau_1}{16\tau_2^{6}(2\pi^2\alpha')^5}\left[\vartheta_3\left(0,\frac{i\tilde{\beta}^2}{\tau_2}\right)-\vartheta_4\left(0,\frac{i\tilde{\beta}^2}{\tau_2}\right)\right]\frac{\Theta_{E_8 \oplus E_8}(-\overline{\tau})}{\vartheta_4(0,2\tau)^{8}\eta(-\overline{\tau})^{24}}.
\end{equation}
where we denoted $\tilde{\beta}^2 = \frac{\beta^2}{4\pi^2\alpha'}$. \\
The large $\tau_2$ limit is again deferred to appendix \ref{appheterotic}. The result is
\begin{equation}
F = -2 V_9\iint_{\mathcal{A}} \frac{d\tau_2d\tau_1 }{2\tau_2(4\pi^2\alpha'\tau_2)^5}e^{-\frac{\beta^2}{4\pi\alpha'}\frac{\tau_1^2+\tau_2^2}{\tau_2}}e^{\pi i \tau_1} e^{3\pi \tau_2}.
\end{equation}
We clearly see that the correction is given by $e^{\pi i \tau_1} e^{3\pi \tau_2}$. The second factor looks like a Hagedorn-like correction, but there is an extra $e^{\pi i \tau_1}$ factor. Intuitively, the heterotic string is left-right asymmetric, so it is not totally unexpected to find such a factor. Factors of this type are also seen in other limits \cite{Alvarez:1986sj}, where they lead to (unexpected) infrared convergence. \\
If we would ignore this $\tau_1$ correction, we would have the prediction that the Hagedorn temperature is equal to 
\begin{equation}
\beta_H = \sqrt{12}\pi\sqrt{\alpha'}
\end{equation}
which is the value predicted by \cite{Alvarez:1986sj}. However, this value is inconsistent with other values in the literature \cite{Gross:1985fr} and with the expected `thermal duality' of the heterotic string \cite{O'Brien:1987pn}. \\
If we proceed correctly and integrate over $\tau_1$, we get
\begin{equation}
\int_{-\infty}^{+\infty}d\tau_1\exp\left(\frac{\beta^2}{4\pi\alpha'\tau_2}\tau_1^2 + \pi i \tau_1\right) = \sqrt{\frac{4\pi^2\alpha'\tau_2}{\beta^2}}\exp\left(-\frac{\pi^3\alpha'}{\beta^2}\tau_2\right).
\end{equation}
We see that we get a correction to the Hagedorn temperature from this integration. \\
To ensure convergence in the large $\tau_2$ limit, we should have
\begin{equation}
\frac{\beta^2}{4\pi\alpha'} + \frac{\pi^3\alpha'}{\beta^2} \geq 3\pi. 
\end{equation}
At first sight, it seems rather strange to have a $\beta$ factor in the denominator, but we now argue that indeed this must be the case.\\
Solving the previous equation gives \emph{two} critical temperatures
\begin{eqnarray}
\beta_{H1} = (2+\sqrt{2})\pi\sqrt{\alpha'},\quad T_{H1} = \frac{1}{\pi\sqrt{\alpha'}}\left(1-\frac{1}{\sqrt{2}}\right), \\
\beta_{H2} = (2-\sqrt{2})\pi\sqrt{\alpha'},\quad T_{H2} = \frac{1}{\pi\sqrt{\alpha'}}\left(1+\frac{1}{\sqrt{2}}\right).
\end{eqnarray}
The partition function converges for $T \leq T_{H1}$ or $T \geq T_{H2}$. The first temperature is the physical one we encounter in heating up a gas of strings. It is necessary that we have two solutions since the heterotic string has a duality symmetry \cite{O'Brien:1987pn} under 
\begin{equation}
\beta \leftrightarrow \frac{2\pi^2\alpha'}{\beta},
\end{equation}
and the previous convergence condition indeed has this symmetry.

\section{Flat spacetime corrections}
\label{corrections}
In the previous sections, we retained only the lowest Fourier mode in the path integral. We saw that we missed certain contributions and we corrected for them by comparing with the known results in flat spacetime. These correction terms obviously have to correspond to the higher Fourier modes that we neglected. In this section we explicitly show that this is indeed the case for strings in flat spacetime. We will determine the exact correction term in the particle path integral result starting from the string path integral alone and by relating the latter to those for particles in uniform magnetic fields. One can presumably also obtain these correction terms by using a double Fourier mode decomposition and regulating several infinite products via zeta-regularization \cite{Alvarez:1985fw}\cite{Alvarez:1986sj}. \\

The starting point is the string action after the modular transformation
\begin{eqnarray}
S = \frac{1}{4\pi\alpha'}\left[\left(1+\frac{\tau_1^2}{\tau_2^2}\right)\int_{0}^{1/\tau_2}d\sigma\int_{0}^{1}d\tau G_{\mu\nu}\partial_{\sigma}X^{\mu}\partial_{\sigma}X^{\nu}\right. \nonumber \\
\left. + 2\frac{\tau_1}{\tau_2}\int_{0}^{1/\tau_2}d\sigma\int_{0}^{1}d\tau G_{\mu\nu}\partial_{\sigma}X^{\mu}\partial_{\tau}X^{\nu}+\int_{0}^{1/\tau_2}d\sigma\int_{0}^{1}d\tau G_{\mu\nu}\partial_{\tau}X^{\mu}\partial_{\tau}X^{\nu}\right].
\end{eqnarray}
In the previous derivation we only kept the lowest worldsheet mode in $\tau$. The expansion of the target coordinate fields is given by
\begin{eqnarray}
X^{i}(\sigma,\tau) =& \sum_{n=-\infty}^{+\infty}e^{2\pi i n \tau_2 \sigma}X^{i}_{n}(\tau), \quad i=1\hdots D-1,\nonumber\\
X^{0}(\sigma,\tau) =& \pm \beta \tau_2 \sigma + \sum_{n=-\infty}^{+\infty}e^{2\pi i n \tau_2 \sigma}X^{0}_{n}(\tau).
\end{eqnarray}
If we keep all of these terms and plug it into the action, we get a total action (in flat space) given by
\begin{equation}
S = S_{0} + \sum_{n=1}^{+\infty}S_{n},
\end{equation}
where the $S_{n}$ are particle actions that combine the modes $\pm n$ and are given by\footnote{A sum over $\mu$ is implied; the metric is flat Euclidean space here.}
\begin{equation}
S_{n} = \frac{1}{4\pi\alpha'}\int_{0}^{1}\frac{d\tau}{\tau_2}\left[2\dot{X}^{\mu}_{n}\dot{X}^{\mu}_{-n} + 4\pi i \tau_1n\left(X^{\mu}_{n}\dot{X}^{\mu}_{-n}-\dot{X}^{\mu}_{n}X^{\mu}_{-n}\right) + 8\pi^2n^2\left(\tau_1^2+\tau_2^2\right)X^{\mu}_{n}X^{\mu}_{-n}\right].
\end{equation}
To see this, note that the integral over $\sigma$ is only non-zero if the two contributing factors have opposite $n$.
The reality of the target fields requires that $X^{\mu}_{-n} = X^{\mu*}_{n}$. Setting
\begin{equation}
X^{\mu}_{n} = A_{n} + i B_{n}, \quad X^{\mu*}_{n} = A_{n} - i B_{n}
\end{equation}
for two real scalar fields $A_{n}$ and $B_{n}$, gives
\begin{align}
S_{n} = \frac{1}{4\pi\alpha'}\int_{0}^{1}\frac{d\tau}{\tau_2}\left[2\dot{A}_{n}\dot{A}_{n} + 2\dot{B}_{n}\dot{B}_{n} + 8\pi\tau_1n\left(A_{n}\dot{B}_{n}-\dot{A}_{n}B_{n}\right) \right. \nonumber \\
\left. + 8\pi^2n^2\left(\tau_1^2+\tau_2^2\right)\left(A_{n}^{2}+B_{n}^2\right)\right]
\end{align}
for one target space component field $X^{\mu}$. 
These fields, that only depend on time, need to be path-integrated over the entire two-dimensional plane (in field space). 
We notice that:
\begin{enumerate}
\item{the corrections are independent of the $n=0$ contribution, so they will generate a term independent of the $X_{0}^{i}$ and $X_{0}^{0}$ fields,}
\item{in the end we will multiply this action by 26 (or 10 times for the superstring) since each target space coordinate $X^{\mu}$ yields the same action,}
\item{the actions corresponding to different $n$ are also decoupled and of identical form, so if we can compute the path integral for one of them, we can compute all of them.}
\end{enumerate}
The way to solve the path integral corresponding to $S_n$ is to reinterpret it as a path integral for a 2D particle moving in a harmonic oscillator potential and interacting with a (imaginary) magnetic field. Exact results are known for such systems \cite{Khandekar:1986ib}\cite{Jones:1971kk}. 
We rename $A_{n} = x$ and $B_{n} = y$. \\
The Hamiltonian corresponding to the Lagrangian above is given by (see e.g. \cite{Gao-Feng:2008})
\begin{equation}
H = \frac{1}{2m}\left[\left(p_{x}+\frac{qBy}{2}\right)^2 + \left(p_{y}-\frac{qBx}{2}\right)^2 + \frac{1}{2}m\omega^2\left(x^2+y^2\right)\right]
\end{equation}
where 
\begin{equation}
m = \frac{1}{\pi \alpha' \tau_2}, \quad \frac{qB}{2} = i\frac{2\tau_1 n}{\alpha' \tau_2}, \quad \omega = \sqrt{4\pi^2n^2\left(\tau_1^2+\tau_2^2\right)}.
\end{equation}
The eigenstates of this Hamiltonian are known exactly:
\begin{equation}
\psi_{N,m_{\ell}}(x,y) \propto \rho^{\left|m_{\ell}\right|}F\left(-N,\left|m_{\ell}\right|+1,\gamma^2\rho^2\right)e^{-\gamma^2\rho^2/2}e^{im_{\ell}\phi}
\end{equation}
where $N = 0,1,2, \hdots$ and $m_{\ell} = 0, \pm1, \pm2, \hdots$ and $\gamma^2 = \frac{qB}{2}$. $F$ is the confluent hypergeometric function and $\rho^2 = x^2+y^2$.
The energies are given by
\begin{equation}
\label{geomsum}
E_{N,m_{\ell}} = \omega_{0}\left(2N+\left|m_{\ell}\right|+1\right)+m_{\ell}\frac{qB}{2m}, \quad \omega_0 = 2\pi n \tau_2.
\end{equation} 
In a general quantum system the heat kernel (path integral of the Euclidean action corresponding to this classical Hamiltonian) is given by
\begin{equation}
K(\textbf{r},t|\textbf{r},0) = \sum_{a}\left|\psi_{a}(\textbf{r})\right|^2e^{-E_{a}t}.
\end{equation}
Integrating the heat kernel over initial and final coordinate gives for normalized eigenfunctions
\begin{equation}
\iint dxdy K(\boldsymbol{\rho},1|\boldsymbol{\rho},0) = \sum_{a}e^{-E_{a}t} \to e^{-2\pi n \tau_2}.
\end{equation}
where we used the fact that in the large $\tau_2$ limit the values $N = m_{\ell} = 0$ dominate. \\

As a check, one can obtain the same result from the known heat kernel \cite{Jones:1971kk}. The relevant heat kernel is given by
\begin{eqnarray}
K(\boldsymbol{\rho},1|\boldsymbol{\rho},0) = \frac{m\omega_0}{2\pi \sinh(\omega_0)}\exp -\left\{\frac{m\omega_0}{\sinh(\omega_0)}(x^2+y^2)\left[\cosh(\omega_0)-\cosh\left(\frac{qB}{2m}\right)\right]\right\}.
\end{eqnarray}
As in the previous derivation, we need to integrate this over $x$ and $y$. These integrals are simple Gaussians and we obtain
\begin{equation}
\label{heeat}
\iint dxdy K(\boldsymbol{\rho},1|\boldsymbol{\rho},0)
= \frac{1}{2\left[\cosh(\omega_0)-\cosh\left(\frac{qB}{2m}\right)\right]}.
\end{equation}
Now we take the $\tau_2 \to \infty$ limit. Since $B$ is purely imaginary, the second term in the denominator has modulus bounded by one, and the first one becomes arbitrarily large. In the limit, the second term is dropped and we finally obtain
\begin{equation}
\iint dxdy K(\boldsymbol{\rho},1|\boldsymbol{\rho},0) \to \frac{1}{2\cosh(\omega_0)} \to e^{-\omega_0} = e^{-2\pi n \tau_2},
\end{equation} 
in agreement with the previous result. Incidentally, one readily checks that (\ref{heeat}) agrees with $\sum_{N,m_{\ell}}e^{-E_{N,m_{\ell}}}$ with energies (\ref{geomsum}). \\

To get all corrections, this has to be taken to the $26^{\text{th}}$ power (the contribution of all target space fields $X^{\mu}$) and we have to take the product of all values of $n$ going from 1 to infinity.
So in all, we get
\begin{equation}
\prod_{j=1}^{26}\prod_{n=1}^{+\infty} e^{-2\pi n \tau_2} = e^{-52\pi \sum_{n=1}^{+\infty}n \tau_2} = e^{52\pi \frac{1}{12} \tau_2}
\end{equation}
where we used $\sum_{n=1}^{+\infty}n = -\frac{1}{12}$. This last step is the analogue of the zeta-regularization used in \cite{Alvarez:1985fw}\cite{Alvarez:1986sj}. \\
The only thing left to do is to include the contribution from the $bc$ ghosts. This is simply the ghost path integral on the torus worldsheet \cite{Polchinski:1998rq} and in the limit $\tau_2 \to \infty$ (after the modular transformation), this gives a factor 
\begin{equation}
e^{-8\pi\tau_2/24}.
\end{equation}
Combining everything finally gives
\begin{equation}
e^{4\pi\tau_2}
\end{equation}
which is precisely the result we got in equation (\ref{bosAdd}) when comparing with the exact result. Note that this derivation is purely from the path integral and never used the quantized string spectrum.\\
We have succeeded in determining the exact correction term for the flat spacetime bosonic string. The question immediately arises whether we can do this also for different backgrounds and for other types of strings. 

\subsection*{Non-trivial background corrections}
As an example of a non-trivial background, consider Rindler spacetime with metric
\begin{equation}
ds^2 = \left(\frac{\rho^2}{\alpha'}\right)d\tau^2 + d\rho^2 + d\mathbf{x}^2_{\perp}.
\end{equation}
In this spacetime we see that $G_{00}$ is quadratic in the field $\rho$, so the corrections are up to fourth order in the fields. Such particle path integrals are in general not exactly calculable. Also, there is a non-trivial mixing between \emph{all} Fourier modes, so we cannot integrate them out one by one. \\
We conclude that the previous tricks (most likely) only work in flat spacetime.

\subsection*{Type II Superstring corrections}

For type II superstrings, we have extra contributions from the worldsheet fermions. The fermions (and superghosts) we need, give in total the following exact contribution to the partition function\footnote{This corresponds to only one type of spin structure on the torus worldsheet. The reason for this is the thermal boundary conditions as discussed in \cite{Alvarez:1986sj}.} \cite{Alvarez:1986sj}
\begin{equation}
2^8\prod_{n=1}^{\infty}(1+q^n)^8(1+\overline{q}^n)^8e^{-4\pi\tau_2/3}
\end{equation}
where $q=\exp(2\pi i \tau)$.
Doing the modular transformation and taking $\tau_2 \to \infty$ limit yields
\begin{equation}
e^{2\pi\tau_2}e^{-\frac{8\pi\tau_2}{6}}.
\end{equation}
The bosons (and the ghosts) give the following contribution
\begin{equation}
e^{\frac{8\pi\tau_2}{6}}.
\end{equation}
We see that we end up with $e^{2\pi\tau_2}$ which is indeed the contribution we identified in section \ref{supers} by comparing with the known flat space result. Mutatis mutandis one can also see that everything works out for the heterotic string.

\section{Alternative approach: the field theory action at one loop}
\label{alternative}
So far, we have obtained a one-loop result that gives the contribution of the winding tachyon\footnote{In what follows, we will call this state (winding number $\pm1$, no discrete momentum and no oscillators) the winding tachyon, even though strictly speaking it is not tachyonic in the regime we are interested in.} to the partition function. All the previous manipulations ensured we only got this contribution and not the oscillators or other quantum numbers. There is however also another way to get the contribution for only this string state, namely the spacetime action. \\
To describe this, let us first take a worldsheet CFT point of view. Consider the one-loop partition function
\begin{equation}
Z(\tau) = \text{Tr}\left(q^{L_0-c/24}\bar{q}^{\bar{L}_0 -c/24}\right) = (q\bar{q})^{-c/24}\text{Tr}\left[e^{2\pi i \tau_1(L_0 - \bar{L}_0) -2\pi\tau_2(L_0 +\bar{L}_0)}\right].
\end{equation}
We are interested in a CFT state that dominates the above partition function as $\tau_2 \to \infty$ (where $\tau$ is living on the modular fundamental domain). This is the state with lowest $L_0 + \bar{L}_0$. To describe this in terms of field theory, we are hence interested in the `geometrization' (i.e. writing in terms of differential operators) of the string Hamiltonian $L_0 + \bar{L}_0$, along the lines of \cite{Dijkgraaf:1991ba}. In Lorentzian signature flat space for instance, this reduces to the Klein-Gordon operator with plane wave solutions $\sim e^{ipx}$. This operator describes how CFT fluctuations propagate in a background without interacting with any other fluctuations. In our case, the states appearing in the above partition function are those living on the thermal manifold and the state that dominates is the thermal scalar. 
This thermal scalar field theory action (which at lowest order in $\alpha'$ coincides with the lowest order \emph{effective} action), can be used to calculate the one-loop contribution to the free energy and should give us the critical behavior of the string gas. The path integral derivation presented in section \ref{deriv} should coincide with the field theory derivation if the large $\tau_2$ limit is correctly taken, that is by integrating out higher Fourier modes instead of setting these to zero.\footnote{As mentioned earlier, so far we have not been able to do this in general.} We will now check to what extent this story is true.\footnote{The reader might object at this point since in this section we consider winding in the modular fundamental domain whereas in the previous sections we discussed `winding' in the modular strip domain. This is however precisely the thermal scalar interpretation that was found long ago \cite{Atick:1988si}: the divergence in the modular strip is reflected in the masslessness of the winding tachyon in the fundamental domain.} \\
Since we could not find a derivation of the relevant action in the literature (and since we will utilize extensions of this derivation several times further on), we first (re)derive the winding tachyon action to lowest order in $\alpha'$ from a spacetime dimensional reduction.

\subsection{Dimensional reduction}
\label{dimred}
The bosonic closed string tachyon action to lowest order in $\alpha'$, is given by
\begin{equation}
S = \frac{1}{2}\int d^{D}x \sqrt{G}e^{-2\Phi}\left(G^{\mu\nu}\partial_{\mu}T\partial_{\nu}T + m^2T^2\right),
\end{equation}
where $T$ is a real scalar field and $\Phi$ is the dilaton field. This also holds for the closed superstring tachyon \emph{before} the GSO projection. \\
Assume now that $x^{0} \sim x^{0}+2\pi R$ while the metric does not depend on $x^{0}$ and has no components $G_{i0}$. We expand the field $T$ in Fourier modes 
\begin{equation}
T(x^{0},x^{i}) = \sum_{n \in \mathbb{Z}}T_{n}(x^{i})e^{\frac{inx^{0}}{R}},
\end{equation}
where $R = \frac{\beta}{2\pi}$.
Plugging this in the tachyon action gives for $T_{n}$
\begin{equation}
S = \pi R \int d^{D-1}x \sqrt{G_{ij}}\sqrt{G_{00}}e^{-2\Phi}\left(G^{ij}\partial_{i}T_{n}\partial_{j}T_{-n}+\frac{k^2G^{00}}{R^2}T_{n}T_{-n}+m^2T_{n}T_{-n}\right).
\end{equation}
Using the reality of $T$ $\left(T_{n} = T^{*}_{-n}\right)$, gives the action
\begin{equation}
S = \pi R \int d^{D-1}x \sqrt{G_{ij}}\sqrt{G_{00}}e^{-2\Phi}\left(G^{ij}\partial_{i}T_{n}\partial_{j}T_{n}^{*}+\frac{k^2G^{00}}{R^2}T_{n}T_{n}^{*}+m^2T_{n}T_{n}^{*}\right),
\end{equation}
and we can restrict to positive $n$. The complex field $T_{n}$ combines both $\pm n$ contributions. \\
This gives the momentum states of the tachyon field directly in the field theory action. The full action contains both winding and momentum fields. When dimensionally reducing the action, we obtain only the momentum states. But T-duality should still be present in the full field action. We thus exploit this and use a T-duality on the action with
\begin{align}
G_{00} &\to \frac{1}{G_{00}}, \\
\Phi &\to \Phi - \frac{1}{2}\ln\left(G_{00}\right), \\
T_{n} &\to T_{w}.
\end{align}
The momentum tachyon field is transformed to a winding tachyon. Also $\sqrt{G}e^{-2\Phi}$ is T-invariant. Thus we arrive at (also using $R \to \alpha'/R$)
\begin{equation}
\label{tachact}
S \sim \int d^{D-1}x \sqrt{G_{ij}}\sqrt{G_{00}}e^{-2\Phi}\left(G^{ij}\partial_{i}T_{w}\partial_{j}T_{w}^{*}+\frac{w^2R^2G_{00}}{\alpha'^2}T_{w}T_{w}^{*}+m^2T_{w}T_{w}^{*}\right).
\end{equation}
This is the tachyon action for a winding $w$ state \cite{Horowitz:1997jc}. \\
For instance for the type II superstring in polar coordinates $G^{00} = \alpha'/\rho^2$, we arrive at $m^{2} \to m^{2} + \frac{w^2\rho^2R^2}{\alpha'^3}$. With $R= \sqrt{\alpha'}$, we get the action used by the authors of \cite{Kutasov:2005rr}\cite{Giveon:2012kp} ($\alpha'=2$)
\begin{equation}
S \sim \int d^{D-1}x \sqrt{G_{ij}}\sqrt{G_{00}}e^{-2\Phi}\left(G^{ij}\partial_{i}T_{w}\partial_{j}T_{w}^{*} + \left(-1 + \frac{w^2\rho^2}{4}\right)T_{w}T_{w}^{*}\right).
\end{equation}
So in all, we start with the tachyon action for the tachyon living in the uncompactified spacetime. Dimensional reduction combined with T-duality then yields the lower-dimensional winding tachyon action. We noted that our original tachyon need not respect the GSO projection, it is enough to be there before the GSO projection to cause compactified tachyons (that \emph{do} satisfy the GSO projection) to appear. \\

This works fine for both bosonic and superstrings, but for the heterotic strings we need to be a little more clever.
The heterotic string has a left-moving tachyon in its spectrum ($m^2 = -4/\alpha'$) just as the bosonic string and a right-moving tachyon ($m^2=-2/\alpha'$) just as the superstring. The latter is projected out due to the GSO projection on the right-moving sector and the first cannot match with anything of the same mass on the right-moving side, so this state also does not exist. This is the story behind the tachyon-free heterotic string theories. \\
We know we should neglect GSO for our covering space tachyon action, but what about this left-right asymmetry? Well, it turns out to work just fine if one \emph{averages} both tachyon masses of the covering space tachyon. We can see why from the flat space spectrum (on the NS side of the right-moving sector) (see e.g. \cite{Schulgin:2011zb} for the spectrum)
\begin{eqnarray}
m^2 = -\frac{4}{\alpha'}+\frac{4N^{left}}{\alpha'}+\left(\frac{n}{R}+\frac{wR}{\alpha'}\right)^2, \\
m^2 = -\frac{2}{\alpha'}+\frac{4N^{right}_{NS}}{\alpha'}+\left(\frac{n}{R}-\frac{wR}{\alpha'}\right)^2,
\end{eqnarray}
where the constraint is $N^{right}_{NS} - N^{left} + \frac{1}{2} = nw$. The crucial difference is the term $\frac{1}{2}$, so that the non-oscillator states \emph{must} have both momentum and winding in the compact direction. Choosing no oscillators and averaging yields
\begin{equation}
m^2 = -\frac{3}{\alpha'} +\frac{n^2}{R^2}+\frac{w^2R^2}{\alpha'^2},
\end{equation}
where $nw=1/2$. Setting $ w= \pm1$ requires $n= \pm 1/2$ (with the same sign). For a general static background, we have accordingly
\begin{equation}
m^2_{total} = -\frac{3}{\alpha'} +\frac{n^2}{R^2G_{00}}+\frac{w^2R^2G_{00}}{\alpha'^2}.
\end{equation}
Specifying to the state we are interested in gives finally for the lowest order $\alpha'$ action\footnote{We dropped the subindex $w$ here for notational convenience.}
\begin{equation}
S \sim \int d^{D-1}x \sqrt{G_{ij}}\sqrt{G_{00}}e^{-2\Phi}\left(G^{ij}\partial_{i}T\partial_{j}T^{*}+\frac{1}{4R^2G_{00}}TT^{*} + \frac{R^2G_{00}}{\alpha'^2}TT^{*} - \frac{3}{\alpha'}TT^{*}\right).
\end{equation}
Hence in the above action (\ref{tachact}), we need to add also a discrete momentum contribution and choose the covering space mass correction equal to the average of the bosonic and the superstring tachyon mass. This action was also written down from a scattering amplitude perspective in \cite{Schulgin:2011zb}.\footnote{Note that our derivation of this action is not watertight in this case, but the derivation using scattering amplitudes \cite{Schulgin:2011zb} shows that this action is the correct one.}\\
In general, one can also obtain the same effective action by analyzing scattering amplitudes as we briefly discuss for the bosonic string in appendix \ref{appscat}. \\
For later convenience, let us assemble the different `local' mass terms for the singly wound string in a single function $m_{local}$ as follows
\begin{align}
\label{mlocal}
m_{local}^2 &= -\frac{4}{\alpha'} + \frac{R^2G_{00}}{\alpha'^2}, \quad \text{for bosonic strings}, \\
m_{local}^2 &= -\frac{2}{\alpha'} + \frac{R^2G_{00}}{\alpha'^2}, \quad \text{for type II superstrings}, \\
m_{local}^2 &= -\frac{3}{\alpha'} +\frac{1}{4R^2G_{00}}+\frac{R^2G_{00}}{\alpha'^2}, \quad \text{for heterotic strings}.
\end{align}
The extension to spacetimes with $G_{0i} \neq 0$ is straightforward (at least for bosonic and type II superstrings). We get the (discrete momentum) tachyon action
\begin{align}
S = \pi R \int d^{D-1}x \sqrt{G}e^{-2\Phi}\left(G^{ij}\partial_{i}T_{n}\partial_{j}T_{n}^{*}+\frac{n^2G^{00}}{R^2}T_{n}T_{n}^{*} \right. \nonumber \\
\left. + G^{0i}\frac{in}{R}\left(T_{n}\partial_{i}T_{n}^{*}- T_{n}^{*}\partial_{i}T_{n}\right) + m^2T_{n}T_{n}^{*}\right),
\end{align}
whereas T-duality now gives
\begin{align}
G_{00} \to \frac{1}{G_{00}}&,\quad G_{0i} \to \frac{B_{0i}}{G_{00}}=0,\quad G_{ij} \to G_{ij} - \frac{G_{0i}G_{0j}}{G_{00}}, \nonumber \\
&\Phi \to \Phi - \frac{1}{2}\ln\left(G_{00}\right),\quad T_{n} \to T_{w}.
\end{align}
The new $G_{0i}$ vanishes because $B_{0i}=0$. Also, $G'^{0i} = 0$ because $G'_{0i} = 0$. We have
\begin{equation}
\sqrt{G}e^{-2\Phi} \to \sqrt{G'}e^{-2\Phi'} = \sqrt{G'_{00}}\sqrt{G'_{ij}}e^{-2\Phi}G_{00} = \sqrt{G_{00}}e^{-2\Phi}\sqrt{G_{ij} - \frac{G_{0i}G_{0j}}{G_{00}}}.
\end{equation}
Thus we arrive at
\begin{equation}
S \sim \int d^{D-1}x \sqrt{G_{ij} - \frac{G_{0i}G_{0j}}{G_{00}}}\sqrt{G_{00}}e^{-2\Phi}\left(G'^{ij}\partial_{i}T_{w}\partial_{j}T_{w}^{*}+\frac{w^2R^2G_{00}}{\alpha'^2}T_{w}T_{w}^{*}+m^2T_{w}T_{w}^{*}\right).
\end{equation}
where $G'^{ij}$ is the matrix inverse of $G_{ij} - \frac{G_{0i}G_{0j}}{G_{00}}$. 
We thus see that the only effect of this more general case is the replacement $G_{ij} \to G_{ij} - \frac{G_{0i}G_{0j}}{G_{00}}$. In what follows we will again restrict to the case $G_{0i} = 0$, but we observe that the final result (after a simple substitution) will still hold in the more general case. This case is in nice agreement with the results in section \ref{exten}.

\subsection{The particle path integral of the field theory action}
\label{qm}
Let us compare the one-loop prediction of this action with our previous derivation from section \ref{deriv}. Since we only path integrate over the tachyon (the metric, dilaton and NS-NS fields are backgrounds), the one-loop effective action $\Gamma^{(1)}$ obtained in this way coincides with the free energy of the system (up to a factor of $\beta$). As a reminder, the first quantized stringy picture and second quantized field picture are related to the free energy as follows:
\begin{align}
F_{gas} &= -\frac{1}{\beta}Z_{part}, \\
F_{gas} &= \frac{1}{\beta}\Gamma^{(1)} = -\frac{1}{\beta}\ln\left(Z_{FT}\right),
\end{align}
where $part$ denotes the particle action derived in section \ref{deriv} and $FT$ denotes the field theory of only the winding tachyon. If the string gas indeed can be described by only the thermal scalar, both formalisms should yield the same expression for the free energy of the string gas.\\
We rewrite the one-loop result in a first-quantized way. The action (\ref{tachact}) derived in the previous section takes the following form (we have dropped the $w$ index of $T_{w=1}$ for notational convenience)
\begin{equation}
S =  \int d^{D-1}x \sqrt{G_{ij}}\sqrt{G_{00}}\left[G^{ij}\partial_{i}T\partial_{j}T^{*}+m_{local}^2TT^{*}\right]
\end{equation}
where $i$ runs over all space indices and $G^{ij}$ is a Euclidean metric on the spatial part of the manifold. We remind the reader that $m_{local}$ is a function of spacetime since it contains metric components. Our goal is to remove the $G_{00}$ contribution to the kinetic term `as much as possible' to hopefully reinterpret this as a particle on a curved background described by only the spatial part of the total manifold. We will proceed very carefully in what follows.\\
We first perform a partial integration in the action to distill the inverse propagator:
\begin{equation}
S =  \int d^{D-1}x \sqrt{G_{ij}}\sqrt{G_{00}}T^{*}\left[-\nabla^{2} - G^{ij}\frac{\partial_{j}\sqrt{G_{00}}}{\sqrt{G_{00}}}\partial_{i} + m_{local}^2\right]T.
\end{equation}
The operator $\nabla^2 = G^{ij}\nabla_{i}\partial_{j}$ denotes the covariant Laplacian on the spatial submanifold.
The operator between square brackets is readily seen to be Hermitian with respect to the inner product
\begin{equation}
\left\langle \psi_1 \right.\left|\psi_2\right\rangle = \int d^{D-1}x \sqrt{G_{ij}}\sqrt{G_{00}} \psi_1(x)^{*} \psi_2(x)
\end{equation}
so its eigenfunctions can be chosen orthonormal and its eigenvalues are real. For convenience, let us call the operator $\hat{\mathcal{O}}$. We now choose a basis of such eigenfunctions and expand
\begin{equation}
\label{eigen}
\psi(x) = \sum_{n}a_{n}\psi_{n}(x), \quad \hat{\mathcal{O}}\psi_{n} = \omega_{n}\psi_{n}, \quad \int d^{D-1}x \sqrt{G_{ij}}\sqrt{G_{00}} \psi_{n}(x)^{*} \psi_{m}(x) = \delta_{n,m}.
\end{equation}
The one-loop action is given by the logarithm of the path integral over $T$ with the above action. In the above basis the path integral gives a product of Gaussian integrals over the $a_{n}$, resulting in\footnote{We ignore prefactors, since they will just end up as an additive contribution to the effective action (due to the logarithm) that are independent of the state in the Hilbert space. Also note that we have a \emph{complex} scalar field so we should square the contributions coming from real scalar fields.}
\begin{equation}
\prod_{n}\left(\frac{1}{\omega_{n}}\right) = \text{det}^{-1}\hat{\mathcal{O}}.
\end{equation}
The one-loop action can be written as
\begin{equation}
\Gamma^{(1)} = -\ln \text{det}^{-1}\hat{\mathcal{O}} = \text{Tr} \ln\hat{\mathcal{O}}.
\end{equation}
Now we use the Schwinger proper time representation of the logarithm
\begin{equation}
\ln(a) = -\int_{0}^{+\infty}\frac{dT}{T}\left(e^{-aT} - e^{-T}\right).
\end{equation}
We drop the $-e^{-T}$ term\footnote{This term is proportional to the size of the Hilbert space, just as the prefactors we ignored above.} which gives
\begin{equation}
\Gamma^{(1)} = - \int_{0}^{+\infty}\frac{dT}{T} \text{Tr} e^{-T\left(-\nabla^{2}+m_{local}^2 - G^{ij}\frac{\partial_{j}\sqrt{G_{00}}}{\sqrt{G_{00}}}\partial_{i}\right)}.
\end{equation}
Note that the net effect of starting with a complex instead of a real scalar field is an overall factor $2$. This corresponds in the previous path integral derivation to the sum over the two winding states. \\

To proceed, we notice a delicate point: despite the fact that it looks like we succeeded in removing all $G_{00}$ dependence from the kinetic term, there is still a non-trivial dependence on it. The trace still contains the $\sqrt{G_{00}}$ measure as is shown in the normalization (\ref{eigen}).  
We still want to remove this factor. This can be done by a simple rescaling of the eigenfunctions. 
Let us define new basis vectors
\begin{equation}
\left|\phi_{n}\right\rangle = G_{00}^{1/4}\left|\psi_{n}\right\rangle
\end{equation}
which are by definition normalized as
\begin{equation}
\int d^{D-1}x \sqrt{G_{ij}}\phi_{n}(x)^{*} \phi_{m}(x) = \delta_{n,m}.
\end{equation}
It now readily follows that the $\left|\phi_{n}\right\rangle$ are eigenstates of the operator obtained by pulling the exponential through the $G_{00}^{-1/4}$ factor as follows
\begin{equation}
e^{-T\hat{\mathcal{O}}}G_{00}^{-1/4} = G_{00}^{-1/4}e^{-T\hat{\mathcal{D}}}.
\end{equation}
In coordinate space, the operator $\hat{\mathcal{D}}$ is simply obtained by setting
\begin{equation}
\hat{\mathcal{D}}f(x) = \frac{\hat{\mathcal{O}}\left(G_{00}^{-1/4} f(x)\right)}{G_{00}^{-1/4}},\quad \forall f(x). 
\end{equation}
We can thus rewrite the previous action as
\begin{equation}
\label{effD}
\Gamma^{(1)} = - \int_{0}^{+\infty}\frac{dT}{T} \text{Tr} e^{-T\hat{\mathcal{D}}}
\end{equation}
where now the $\sqrt{G_{00}}$ is removed from the integral measure in the trace (we have in effect changed the definition of the inner product on our Hilbert space).\\
Transforming $\hat{\mathcal{O}}$ in this fashion, we obtain the following operator in the exponential
\begin{equation}
\hat{\mathcal{D}} = -\nabla^{2} -\frac{3}{16}\frac{G^{ij}\partial_iG_{00}\partial_jG_{00}}{G_{00}^2} + \frac{\nabla^2 G_{00}}{4G_{00}}+ m_{local}^2
\end{equation}
where the terms involving only one $\partial_{i}$ have miraculously dropped out. As a check, we see that this operator $\hat{\mathcal{D}}$ is Hermitian with respect to the canonical inner product on the spatial submanifold
\begin{equation}  
\left\langle \phi_1 \right.\left|\phi_2\right\rangle = \int d^{D-1}x \sqrt{G_{ij}}\phi_1(x)^{*}\phi_2(x).
\end{equation}
The only corrections we have are two terms that behave as a potential for the particle. 
We denote these for convenience in what follows as $K(x)$:
\begin{equation}
\label{K}
K(x) =-\frac{3}{16}\frac{G^{ij}\partial_iG_{00}\partial_jG_{00}}{G_{00}^2} + \frac{\nabla^2 G_{00}}{4G_{00}}.
\end{equation}
The exponential in (\ref{effD}) needs to be given a Lagrangian interpretation. 
So we seek a path integral description for a system with Hamiltonian
\begin{equation}
H = p_i p_j G^{ij} + m_{local}^2 + K(x).
\end{equation}
The corresponding (Euclidean) Lagrangian is given by
\begin{equation}
L_E = \frac{1}{4}\dot{x}^{i}\dot{x}^{j}G_{ij} + m_{local}^2 + K(x).
\end{equation}
The last step is then to give a path integral representation for $e^{-TH}$ which finally results in\footnote{The appearance of the $\sqrt{G_{ij}}$ in the measure is natural from coordinate invariance. It can also be explicitly derived (see \cite{Abers:1973qs} for the particle case and \cite{Weinberg:1995mt} (chapter 9) for the field theory case).}
\begin{equation}
\Gamma^{(1)} = - \int_{0}^{+\infty}\frac{dT}{T}\int_{S^{1}} \left[\mathcal{D}x\right]\sqrt{G_{ij}}e^{-\int_{0}^{T}dt\left(\frac{1}{4}G_{ij}(x)\dot{x}^{i}\dot{x}^{j} +  m_{local}^2 + K(x)\right)}.
\end{equation}
The $S^{1}$ denotes periodic boundary conditions on the path: $x(0)=x(T)$.
Performing the substitution $t \to \pi \alpha' t$ gives the path integral
\begin{equation}
\label{FTaction}
\Gamma^{(1)} = - \int_{0}^{+\infty}\frac{dT}{T}\int_{S^{1}} \left[\mathcal{D}x\right]\sqrt{G_{ij}}e^{-\frac{1}{4\pi\alpha'}\int_{0}^{T}dt\left(G_{ij}(x)\dot{x}^{i}\dot{x}^{j} + 4\pi^2\alpha'^2\left(m_{local}^2 + K(x\right)\right)}.
\end{equation}
For instance, filling in the correct value for $m_{local}$ (\ref{mlocal}) when the bosonic flat space Hagedorn temperature is to be used, gives
\begin{equation}
\label{partaction}
\Gamma^{(1)} = - \int_{0}^{+\infty}\frac{dT}{T}\int_{S^{1}} \left[\mathcal{D}x\right]\sqrt{G_{ij}}e^{-\frac{1}{4\pi\alpha'}\int_{0}^{T}dt\left(G_{ij}(x)\dot{x}^{i}\dot{x}^{j} + 4\pi^2\alpha'^2\left(-\frac{4}{\alpha'} + \frac{R^2G_{00}}{\alpha'^2} + K(x)\right)\right)}.
\end{equation}
\\
We can compare this with the result from section \ref{deriv}, when we identify $\Gamma^{(1)}$ directly with $\beta F$. For convenience, we rewrite the result given there:
\begin{equation}
\label{Firstaction} 
\beta F = -\int_0^{+\infty} \frac{d \tau_2}{\tau_2} \int_{S^{1}} \left[ \mathcal{D}X \right] \sqrt{G_{ij}} \exp -\frac{1}{4\pi \alpha'}\left[ \int_0^{\tau_2} dt \left( G_{ij} \partial_t X^i \partial_t X^j + \beta^2  G_{00} - \beta_H^2 \right) \right]. 
\end{equation}
Translating (\ref{FTaction}) to this notation, we find
\begin{equation}
\label{pint} 
\beta F = - \int_{0}^{+\infty}\frac{d\tau_2}{\tau_2}\int_{S^{1}} \left[\mathcal{D}X\right]\sqrt{G_{ij}}\exp -S
\end{equation}
where
\begin{equation}
\label{FFTaction}
S = \frac{1}{4\pi\alpha'} \left[\int_{0}^{\tau_2} dt \left\{G_{ij}\partial_t X^{i}\partial_ tX^{j} + 4\pi^2\alpha'^2\left(m_{local}^2 + K(X)\right)\right\}\right].
\end{equation}

Now let us compare the second quantized field theory result (\ref{FFTaction}) with the first quantized result (\ref{Firstaction}).

\begin{itemize}
\item{The particle action (\ref{FFTaction}) naturally lives in one dimension less than the original problem. Here this occurs due to the dimensional reduction used to arrive at the field theory action. In (\ref{Firstaction}), this happened because we integrated out the Euclidean time coordinate explicitly.}
\item{
For bosonic and type II superstrings we have the equality\footnote{Using the appropriate value of $\beta_{H}$.}
\begin{equation}
4\pi^2\alpha'^2m_{local}^2 = \beta^2G_{00} - \beta_{H}^2.
\end{equation}
For the heterotic string, the story changes a bit. The mass term is now of the form
\begin{equation}
\label{massheter}
4\pi^2\alpha'^2m_{local}^2 = -\beta_{H}^2 + \beta^2G_{00} + \frac{\pi^2\alpha'^2G^{00}}{R^2}.
\end{equation}
Note that $\beta_{H}^2 = 12\pi^2\alpha'^2$. This corresponds to the averaging of the bosonic and type II tachyon masses as we argued in section \ref{dimred}. The reader should not be confused at this point: this $\beta_{H}$ is not equal to the Hagedorn temperature, not even in the flat space. It is simply the mass of the covering space tachyon. The final term in (\ref{massheter}) is not found using the naive path integral result and should be added (just like the $\beta_{H}^2$-term). In flat spacetime, we found that this correction actually originated from a $\tau_1$-dependent contribution as we discussed in section \ref{heterotic}.
}
\item{The second quantized result (\ref{FFTaction}) has an extra term (denoted as $K(x)$). 
This suggests we missed this term in the derivation in section \ref{deriv} in the `worldsheet dimensional reduction', which should be incorporated in the result in the same way as the $\beta_{H}^2$ contribution. This term alters the random walk behavior discussed in \cite{Kruczenski:2005pj} and we will show in an upcoming paper \cite{Mertens:2013zya}\cite{examples} that it gives crucial modifications. This extra term disappears when choosing a flat metric, so in the flat case we have perfect agreement between the two approaches. The extra term cannot in general be discarded in any approximation since it is not a higher order curvature contribution (we will discuss this more extensively in \cite{Mertens:2013zya}\cite{examples}).}

\item{This section only focused on the case where there is only a background metric. We can extend this result to include a background NS-NS field. This is discussed in appendix \ref{KR}. The upshot is that we get precisely the result from section \ref{exten} but with another extra correction term.}

\end{itemize}

\subsection{General discussion of corrections to the particle action}
Let us look in general to the corrections of the particle action that we derived in section \ref{deriv}. In what follows, we define a correction term as a term that is missed in the naive worldsheet dimensional reduction for $\tau_2 \to \infty$ discussed in section \ref{deriv}. \\

Firstly, we have a correction term proportional to the mass of the most tachyonic mode in the \emph{cover} of the manifold that `unwraps' the thermal direction. Note that this only depends on the \emph{type} of string theory used and not on the manifold itself. For bosonic and type II superstrings, this equals the (flat space) Hagedorn correction to the action, but this need not be precisely the Hagedorn temperature of the space. As an example of the latter case, consider the WZW $AdS_3$ bosonic string background (we will discuss this model in a companion paper \cite{examples}). The correction is in this case the Hagedorn temperature of the flat space bosonic string, but this is not equal to the $AdS_3$ Hagedorn temperature.\footnote{The latter temperature is \emph{larger} than the flat space bosonic Hagedorn temperature.} We conclude that this term represents strings that \emph{locally} approach the Hagedorn temperature, but this need not give a \emph{global} divergence in the free energy. This is very reminiscent of the well-known fact that negative mass$^2$ particles in $AdS$ spacetimes can be stable if their mass$^2$ is not too low (the Breitenlohner-Freedman bound). \\
For heterotic strings, we found an additional correction that gives the discrete momentum contribution of the winding tachyon.\\

In flat spacetime, this is the only correction. The one-loop contribution is a simple sum over all string states without including interactions. When restricting to the winding $\pm 1$ states, this is twice the field theory vacuum bubble diagram. We saw that we have an exact matching between our path integral result and the effective action calculated from the thermal scalar field theory action.\\
 
Secondly, we found a correction term $K(x)$ (\ref{K}) obtained from a non-trivial $G_{00}$ metric component. This term alters the random walk behavior and there is no rationale in neglecting it.\\

This need not be the end of the story however, as $\alpha'$ corrections to the spacetime action might be important. Their influence and appearance is rather subtle. The precise effect of these higher order $\alpha'$ corrections in general on the random walk behavior is something we are still further investigating. In the case of Rindler spacetime, these corrections are understood in full detail as we show in \cite{Mertens:2013zya}. Other examples with and without $\alpha'$ corrections will be discussed elsewhere \cite{examples}. 

\section{Conclusion}
We have reviewed and extended the path integral derivation of the random walk behavior of near-Hagedorn thermodynamics \cite{Kruczenski:2005pj}. We analyzed this particle path integral for several string types. The worldsheet dimensional reduction misses some correction terms that we cannot (yet) determine solely from the path integral. In flat spacetime, we calculated these explicitly using known particle path integrals. We then changed gears and calculated the one-loop contribution of the field theory action of the dominant winding string. 
These two results should coincide since both methods focus precisely on the same string state. We checked this and it works out well for flat space. 
For curved backgrounds however, we found a discrepancy: the spacetime action has an extra term in the action. We identified this as something we missed in the (naive) worldsheet dimensional reduction. This identifies both approaches. Both approaches (worldsheet and spacetime) have their advantages though the worldsheet path integral approach has a direct connection to the random walk picture of string thermodynamics in the microcanonical ensemble. However, this approach is computationally more challenging because the $\tau_2 \to \infty$ limit involves coupling to higher Fourier modes which we drop in a first approximation. These are however important to attain the full result in curved space. We found that the correction terms can be divided in three different types:
\begin{itemize}
\item{Firstly we have a correction that simply introduces the mass of the flat space tachyon state. We found these explicitly in section \ref{deriv} for the bosonic string and in sections \ref{supers} and \ref{heterotic} for type II and heterotic strings. For heterotic strings however, we found that we have to introduce a second term as well (we discussed this in section \ref{dimred}), corresponding to the discrete momentum quantum number of the winding tachyon.}
\item{Secondly we have a correction coming from the $G_{00}$ component. We found this by comparing the string path integral with the effective action of the winding tachyon. This term was explicity determined in equation (\ref{K}).}
\item{Other $\alpha'$ correction terms might appear (originating from the field theory action), but we postpone their treatment to other work \cite{Mertens:2013zya}\cite{examples}.}
\end{itemize}
We conclude that the random walk picture is modified due to all these correction terms. 
Several questions arise in this process: is there really a winding mode in the string spectrum, especially if the space does not topologically support winding modes? Can we get a handle on the higher correction terms? In \cite{Mertens:2013zya} we answer these questions for black holes, when we take a near-horizon Rindler approximation. We will find precisely which correction terms are required to have a full description of the near-Hagedorn critical behavior of a string gas surrounding a black hole. With this random walk description and the interpretation as a long highly excited string, we will obtain a realization of Susskind's idea of long strings surrounding black hole horizons. We also study several other specific examples of these methods in a companion paper \cite{examples}.

\section*{Acknowledgements}
The authors would like to thank Ben Craps, David Dudal and Lihui Liu for several valuable discussions and Chung-I Tan for useful correspondence. We also thank David Dudal for a careful reading of the manuscript. TM thanks the UGent Special Research Fund for financial support.

\appendix

\section{Extension path integral to non-static spacetimes}
\label{AppAA}
We start with the action (\ref{particleaction}) in the general case $G_{0i} \neq 0$:
\begin{equation}
S_{part} =  \frac{1}{4\pi \alpha' } \left[ \beta^2\frac{\left|\tau\right|^2}{\tau_2^2} \int_0^{\tau_2} dt G_{00} - \beta_H^2 \tau_2 \pm 2 \frac{\tau_1}{\tau_2} \beta \int_0^{\tau_2} dt G_{0\mu} \partial_t X^\mu 
+\int_0^{\tau_2} dt G_{\mu\nu} \partial_t X^\mu \partial_t X^\nu \right].
\end{equation}
The classical equation of motion of $X^{0}$ is given by
\begin{equation}
\pm \frac{\tau_1}{\tau_2}\beta \partial_t G_{00} + \partial_t \left(G_{00}\partial_t X^{0}\right) + \partial_t \left(G_{0i}\partial_t X^{i}\right) = 0. 
\end{equation}
Integrating with respect to $t$ gives
\begin{equation}
\pm \frac{\tau_1}{\tau_2}\beta G_{00} + G_{00}\partial_t X^{0} + G_{0i}\partial_t X^{i} = C. 
\end{equation}
Again integrating and using periodicity of $X^{\mu}$ fixes 
\begin{equation}
C = \frac{1}{\left\langle G_{00}^{-1}\right\rangle}\left[\pm \tau_1 \beta + \left\langle \frac{G_{0i}\partial_t X^{i}}{G_{00}}\right\rangle\right].
\end{equation}
One readily finds for the on-shell classical action (including only the $X^{0}$-dependent contributions)
\begin{align}
4\pi\alpha'S_{on-shell} &= \mp 2\frac{\tau_1}{\tau_2}\beta \left\langle G_{0i}\partial_t X^{i}\right\rangle - \frac{\tau_1^2}{\tau_2^2}\beta^2\left\langle G_{00}\right\rangle + \frac{\tau_1^2\beta^2}{\left\langle G_{00}^{-1}\right\rangle} \pm \frac{2\tau_1\beta}{\left\langle G_{00}^{-1}\right\rangle}\left\langle \frac{G_{0i}\partial_t X^{i}}{G_{00}}\right\rangle \nonumber \\
&+ \frac{1}{\left\langle G_{00}^{-1}\right\rangle}\left\langle \frac{G_{0i}\partial_t X^{i}}{G_{00}}\right\rangle^2 - \left\langle \frac{G_{0i}G_{0j}}{G_{00}}\partial_t X^{i}\partial_t X^{j}\right\rangle.
\end{align}
Setting $X^{0} = X^{0,cl} + \tilde{X}^{0}$, one arrives at the following total action
\begin{align}
4\pi\alpha'S_{total} &=  \beta^2\int_0^{\tau_2} dt G_{00} - \beta_H^2 \tau_2 + \frac{\tau_1^2\beta^2}{\left\langle G_{00}^{-1}\right\rangle} \pm \frac{2\tau_1\beta}{\left\langle G_{00}^{-1}\right\rangle}\left\langle \frac{G_{0i}\partial_t X^{i}}{G_{00}}\right\rangle \nonumber \\
&+ \frac{1}{\left\langle G_{00}^{-1}\right\rangle}\left\langle \frac{G_{0i}\partial_t X^{i}}{G_{00}}\right\rangle^2 + \left\langle \left(G_{ij}-\frac{G_{0i}G_{0j}}{G_{00}}\right)\partial_t X^{i}\partial_t X^{j}\right\rangle + \left\langle G_{00} \partial_t \tilde{X}^{0} \partial_t \tilde{X}^{0}\right\rangle.
\end{align}
Finally performing the $\tau_1$ integration, the first term in the second line is precisely cancelled and the $\tilde{X}^{0}$ path integral again cancels the $\left\langle G_{00}^{-1}\right\rangle$ prefactors. Since 
\begin{equation}
\sqrt{G} = \sqrt{G_{00}}\sqrt{G_{ij} - \frac{G_{0i}G_{0j}}{G_{00}}},
\end{equation}
we finally end up with
\begin{equation} 
Z_p = 2\int_0^\infty \frac{d \tau_2}{2\tau_2} \int \left[ d\vec X \right] \sqrt{\prod \det \left(G_{ij} - \frac{G_{0i}G_{0j}}{G_{00}}\right)} \exp - S_p(\vec X) 
\end{equation}
where 
\begin{equation}
S_p = \frac{1}{4\pi \alpha'}\left[ \beta^2 \int_0^{\tau_2} dt G_{00} - \beta_H^2 \tau_2 
+\int_0^{\tau_2} dt \left(G_{ij} - \frac{G_{0i}G_{0j}}{G_{00}}\right) \partial_t X^i \partial_t X^j\right],
\end{equation}
which is the expression shown in section \ref{nonstat}.

\section{\boldmath Large $\tau_2$ limit of several string theories}

\subsection*{Superstring}
\label{appsuper}
The free energy for superstrings is given by the following expression \cite{Alvarez:1986sj}
\begin{equation}
F = -2 V_9\int_{0}^{+\infty}d\tau_2\int_{-1/2}^{1/2}\frac{d\tau_1}{\tau_2^{6}(2\pi^2\alpha')^5}\left[\vartheta_3\left(0,\frac{i\beta^2}{4\pi^2\alpha'\tau_2}\right)-\vartheta_4\left(0,\frac{i\beta^2}{4\pi^2\alpha'\tau_2}\right)\right]\left|\vartheta_4(0,2\tau)\right|^{-16}
\end{equation}
where 
\begin{equation}
\vartheta_3(0,\tau) = \sum_{n=-\infty}^{+\infty}q^{n^2/2}, \quad \vartheta_4(0,\tau) = \sum_{n=-\infty}^{+\infty}(-1)^n q^{n^2/2}.
\end{equation}
For the modular functions, we follow the definitions of \cite{Polchinski:1998rq}.
We do a modular transformation $\tau \to -\frac{1}{\tau}$. 
The factor $\frac{d\tau_1d\tau_2}{\tau_2^2}$ is invariant under modular transformations. Also,
\begin{equation}
\vartheta_3\left(0,\frac{i\beta^2}{4\pi^2\alpha'\tau_2}\right)-\vartheta_4\left(0,\frac{i\beta^2}{4\pi^2\alpha'\tau_2}\right) = \sum_{n=-\infty}^{+\infty}(1-(-1)^n)e^{-\frac{\beta^2n^2}{4\pi\alpha'}\frac{\tau_1'^2+\tau_2'^2}{\tau_2'}}.
\end{equation}
From now on we drop the primes.
We can take the $\tau_2 \to +\infty$ limit of the expression:
\begin{equation}
\label{integrand}
F = -2 V_9\iint_{\mathcal{A}} d\tau_2\frac{d\tau_1}{\tau_2^{2}(2\pi^2\alpha')^5}\left(\frac{\left|\tau\right|^2}{\tau_2}\right)^4\left[\sum_{n=-\infty}^{+\infty}(1-(-1)^n)e^{-\frac{\beta^2n^2}{4\pi\alpha'}\frac{\tau_1^2+\tau_2^2}{\tau_2}}\right]\left|\vartheta_4\left(0,-\frac{2}{\tau}\right)\right|^{-16}.
\end{equation}
The $\vartheta_4$ function has the modular property
\begin{equation}
\vartheta_4\left(0,-\frac{1}{\tau}\right) = (-i\tau)^\frac{1}{2}\vartheta_2(0,\tau),
\end{equation}
and the $\vartheta_2$ function has the following product expansion
\begin{equation}
\vartheta_2(0,\tau) = 2e^{\pi i \tau/4} \prod_{m=1}^{+\infty}(1-q^m)(1+q^m)^2.
\end{equation}
We notice that $\left|\vartheta_2\right|$ for $\tau_2 \to \infty$ has no contribution from the infinite product. So
\begin{equation}
\left|\vartheta_2(0,\tau)\right| \to 2e^{-\pi \tau_2 /4}
\end{equation}
irrespective of the value of $\tau_1$. We finally arrive at
\begin{equation}
\left|\vartheta_4\left(0,-\frac{2}{\tau}\right)\right|^{-16} \to \left|\tau\right|^{-8}2^{-8}e^{2\pi\tau_2}.
\end{equation}
The sum in the integrand (\ref{integrand}) is dominated by $n= \pm 1$ in the limit $\tau_2 \to +\infty$. Plugging all this into the integral finally gives
\begin{eqnarray}
F &= -2 V_9\iint_{\mathcal{A}} d\tau_2\frac{d\tau_1}{\tau_2^6(2\pi^2\alpha')^5}\left[4e^{-\frac{\beta^2}{4\pi\alpha'}\frac{\tau_1^2+\tau_2^2}{\tau_2}}\right]2^{-8}e^{2\pi\tau_2} \nonumber \\
&= -2 V_9\iint_{\mathcal{A}} \frac{d\tau_2d\tau_1}{2\tau_2}\frac{1}{(4\pi^2\alpha'\tau_2)^5}e^{-\frac{\beta^2}{4\pi\alpha'}\frac{\tau_1^2+\tau_2^2}{\tau_2}}e^{2\pi\tau_2}.
\end{eqnarray}
which is the result stated in section \ref{supers}.

\subsection*{Heterotic string}
\label{appheterotic}
The free energy for the $E_8 \times E_8$ heterotic string is given by \cite{Alvarez:1986sj}
\begin{align}
F = -2 V_9\int_{0}^{+\infty}d\tau_2\int_{-1/2}^{1/2}\frac{d\tau_1}{16\tau_2^{6}(2\pi^2\alpha')^5}\left[\vartheta_3\left(0,\frac{i\beta^2}{4\pi^2\alpha'\tau_2}\right)-\vartheta_4\left(0,\frac{i\beta^2}{4\pi^2\alpha'\tau_2}\right)\right] \nonumber \\
\times \frac{\Theta_{E_8 \oplus E_8}(-\overline{\tau})}{\vartheta_4(0,2\tau)^{8}\eta(-\overline{\tau})^{24}}.
\end{align}
We again use the modular transformation $\tau \to -1/\tau$ and the limiting behavior of the modular functions (as $\tau_2 \to \infty$)
\begin{eqnarray}
\vartheta_4\left(0,-\frac{2}{\tau}\right)^{-8} \to \frac{e^{-\pi i \tau}}{\tau^42^4}, \\
\eta\left(\frac{1}{\overline{\tau}}\right) = (i\overline{\tau})^{1/2}\eta(-\overline{\tau}) \to (i\overline{\tau})^{1/2} e^{-\frac{\pi i \overline{\tau}}{12}}.
\end{eqnarray}
Using $\Theta_{E_8 \oplus E_8} = \Theta_{E_8}^2$ and $\Theta_{E_8} = \frac{1}{2}\left(\vartheta_2^8+\vartheta_3^8+\vartheta_4^8\right)$, we see that\footnote{This result is more general: given a lattice theta function $\Theta_{\Gamma}(\tau)$, the limit for $\tau_2 \to +\infty$ is always equal to 1. This implies that this derivation also holds for the SO(32) heterotic string (based on the $\Gamma_{16}$ lattice).}
\begin{equation}
\Theta_{E_8 \oplus E_8}\left(\frac{1}{\overline{\tau}}\right) = (-i\overline{\tau})^8 \Theta_{E_8 \oplus E_8}\left(-\overline{\tau}\right) \to (-i\overline{\tau})^8
\end{equation}
where the contributions from $\vartheta_3$ and $\vartheta_4$ give a factor of 4 that cancels the denominator. We now arrive at\footnote{There is a subtlety here: we approximate $\tau_1^2 + \tau_2^2 \approx \tau_2^2$, which is only valid for $\tau_2 \gg \tau_1$. So the $\tau_2 \to \infty$ limit is taken before the integral over $\tau_1$ is performed.} 
\begin{align}
F &= -2 V_9\iint_{\mathcal{A}} \frac{d\tau_2d\tau_1 \tau_2^4}{16\tau_2^2(2\pi^2\alpha')^5}\left[4e^{-\frac{\beta^2}{4\pi\alpha'}\frac{\tau_1^2+\tau_2^2}{\tau_2}}\right]\frac{e^{-\pi i \tau}}{\tau^42^4}(i\overline{\tau})^{-12} e^{2\pi i \overline{\tau}} (-i\overline{\tau})^8 \nonumber \\
&= -2 V_9\iint_{\mathcal{A}} \frac{d\tau_2d\tau_1 }{2\tau_2(4\pi^2\alpha'\tau_2)^5}e^{-\frac{\beta^2}{4\pi\alpha'}\frac{\tau_1^2+\tau_2^2}{\tau_2}}e^{\pi i \tau_1} e^{3\pi \tau_2}.
\end{align}
which is the result stated in section \ref{heterotic}.

\section{Scattering amplitudes}
\label{appscat}
In this appendix we compute the scattering amplitude of one graviton and two winding ($w= \pm1$) tachyons for the bosonic string. We first present the amplitudes as computed from string theory and then we reproduce these amplitudes from the thermal scalar field theory action. To avoid awkwardness, we return to Lorentzian signature and consider the $25$-direction to be compactified with radius $R$. Upon Wick rotating the resulting action, we will see that the spacetime action has the form of the thermal scalar action (\ref{tachact}). For the indices, we will denote $M,N$ as 26-dimensional indices and $\mu,\nu$ as 25-dimensional indices.

\subsection*{Stringy amplitudes}
As the vertex operators we take
\begin{equation}
\frac{g_{c,25}}{\alpha'}:\left(\partial X^{M} \bar{\partial} X^{N} + \partial X^{N} \bar{\partial} X^{M}\right)e^{i k \cdot X}:
\end{equation}
for the graviton and KK scalar ($M=N=25$) and the winding tachyon vertex operators are given by
\begin{equation}
g_{c,25}:e^{ik_{L} \cdot X_{L}(z) + ik_{R} \cdot X_{R}(\overline{z})}:
\end{equation}
where we denoted $g_{c,25} = \frac{g_{c}}{\sqrt{2 \pi R}}$. \\
Following standard arguments \cite{Polchinski:1998rq}, we arrive at the following scattering amplitudes\footnote{Note that this differs slightly from the results in \cite{Polchinski:1998rq} due to a somewhat different normalization of the graviton vertex operators.} (where the two tachyons are wound with $w=1$ and $w=-1$)
\begin{eqnarray}
\label{graviton}
G^{\mu\nu} \to &-\pi i g_{c,25}(2\pi)^{25}\delta^{25}(k_1+k_2+k_3)k^{\mu}_{23}k^{\nu}_{23}, \\
G^{\mu25} \to &-\pi i g_{c,25}(2\pi)^{25}\delta^{25}(k_1+k_2+k_3)k^{\mu}_{23}(k^{25}_{L23}+k^{25}_{R23}), \\
\label{scalar}
G^{2525} \to &-\pi i g_{c,25}(2\pi)^{25}\delta^{25}(k_1+k_2+k_3)k^{25}_{L23}k^{25}_{R23}.
\end{eqnarray}
The last amplitude becomes (using $k^{25}_{L23} = \frac{2R}{\alpha'}$ and $k^{25}_{R23} = -\frac{2R}{\alpha'}$)
\begin{equation}
G^{2525} \to \pi i g_{c,25}(2\pi)^{25}\delta^{25}(k_1+k_2+k_3)\frac{4R^2}{\alpha'^2}.
\end{equation}

\subsection*{Field theory amplitudes}
The field theory action is
\begin{equation}
S = -\int d^{25}x \sqrt{-G_{\mu\nu}}\sqrt{G_{2525}}\left(G^{\mu\nu}\partial_{\mu}T\partial_{\nu}T^{*}+\frac{R^2 G_{2525}}{\alpha'^2}TT^{*}-\frac{4}{\alpha'}TT^{*}\right)
\end{equation}
where we have absorbed $\sqrt{2\pi R}$ in the tachyon field.\footnote{This action is the Lorentzian signature action for a complex tachyon with winding number $w = \pm 1$. Note that we do not include $G_{\mu25}$ dependence, although one can readily generalize the arguments given here to this more general case.} We now show that this action reproduces the graviton and Kaluza-Klein scalar amptitudes determined above. Expanding $G_{MN} = \eta_{MN} - 2\kappa_{25}e_{MN}f(x)$ results in
\begin{equation}
\sqrt{-G_{\mu\nu}}\sqrt{G_{2525}} = \sqrt{\det\left(\eta_{MN} - 2\kappa_{25}e_{MN}f(x)\right)} \approx 1 + \text{Tr}(- 2\kappa_{25}e_{MN}f(x)) =1
\end{equation}
because the polarization tensor is traceless for a graviton. We then expand the action resulting in
\begin{eqnarray}
S &= -\int d^{25}x \left(\eta^{\mu\nu}\partial_{\mu}T\partial_{\nu}T^{*} + 2\kappa_{25} e^{\mu\nu}f(x)\partial_{\mu}T\partial_{\nu}T^{*}  + \frac{R^2}{\alpha'^2}TT^{*} \right. \nonumber \\
&\left.- \frac{R^2 2\kappa_{25} e_{2525}}{\alpha'^2}f(x)TT^{*}-\frac{4}{\alpha'}TT^{*}\right).
\end{eqnarray}
We clearly see a kinetic term for the tachyon (first term) and two mass terms (third and fifth term). The two other terms describe interactions with the graviton. The second term corresponds to graviton-tachyon-tachyon scattering, while the fourth one corresponds to KK scalar-tachyon-tachyon scattering.\\
The (graviton)-(winding tachyon)-(winding tachyon) amplitude becomes
\begin{equation}
A^{\mu\nu} \propto  2i\kappa_{25}e^{\mu\nu}k^{2}_{\mu}k^{3}_{\nu}
= -i\frac{\kappa_{25}}{2}e^{\mu\nu}k^{23}_{\mu}k^{23}_{\nu}.
\end{equation}
When writing $\kappa_{25} = 2\pi g_{c,25}$ and including the kinematic factors, we get
\begin{equation}
\label{graviton2}
A^{\mu\nu} = -i\pi g_{c,25}e^{\mu\nu}k^{23}_{\mu}k^{23}_{\nu}(2\pi)^{25}\delta^{25}(k_1+k_2+k_3).
\end{equation}
The (KK scalar)-(winding tachyon)-(winding tachyon) amplitude becomes
\begin{equation}
A^{2525} \propto 2i\kappa_{25}e_{2525}\frac{R^2}{\alpha'^2} .
\end{equation}
We obtain
\begin{equation}
\label{scalar2}
A^{2525} =  4 i\pi g_{c,25}e^{2525}(2\pi)^{25}\delta^{25}(k_1+k_2+k_3)\frac{R^2}{\alpha'^2}.
\end{equation}
Note that $e^{2525} = e_{2525}$ since indices of $e_{MN}$ are raised and lowered with $\eta_{MN}$. \\
We conclude that the stringy amplitudes (\ref{graviton}) and (\ref{scalar}) agree with the field amplitudes (\ref{graviton2}) and (\ref{scalar2}) respectively.\footnote{If we choose discrete momentum states instead of winding states, this would result in the stringy amplitude with $k^{25}_{L23}k^{25}_{R23} = \frac{n^2}{R^2}$. So in all, this would differ by a minus sign and a factor of $\frac{R^4}{\alpha'^2}$, precisely the difference in the effective action: the minus sign comes from the $G^{2525}$ component (instead of $G_{2525}$) and the other factor comes from the T-duality step to obtain the winding action.} Wick rotating then immediately yields the thermal scalar action (\ref{tachact}). Obviously the above action was not entirely general, for instance the dilaton field or the $G^{\mu25}$ field couplings are not present. One can readily generalize the above to also include these contributions.

\section{Extension of the thermal scalar field theory to a background Kalb-Ramond field}
\label{KR}
We make a final extension to the field theory result and include also a non-zero NS-NS field. We know that to lowest order, the covering-space tachyon action does not couple to the NS-NS field. This however does not imply that there is no influence of the NS-NS background as one can readily check that the scattering amplitudes for a $B_{0i}$ component and two winding tachyons does not vanish. Hence we do expect a coupling. \\
The (discrete momentum) tachyon action for the complex field $T_{n}$ is given by
\begin{align}
S = \pi R \int d^{D-1}x \sqrt{G}e^{-2\Phi}\left(G^{ij}\partial_{i}T_{n}\partial_{j}T_{n}^{*} +\frac{n^2G^{00}}{R^2}T_{n}T_{n}^{*} \right.\nonumber \\
\left. + G^{0i}\frac{in}{R}\left(T_{n}\partial_{i}T_{n}^{*}- T_{n}^{*}\partial_{i}T_{n}\right)+ m^2T_{n}T_{n}^{*}\right),
\end{align}
The T-duality is given by
\begin{align}
G_{00} \to \frac{1}{G_{00}}&,\quad G_{0i} \to \frac{B_{0i}}{G_{00}},\quad G_{ij} \to G_{ij} - \frac{G_{0i}G_{0j}}{G_{00}} + \frac{B_{0i}B_{0j}}{G_{00}}, \nonumber \\
&\Phi \to \Phi - \frac{1}{2}\ln\left(G_{00}\right),\quad T_{n} \to T_{w}.
\end{align}
We have
\begin{equation}
\sqrt{G}e^{-2\Phi} \to \sqrt{G'}e^{-2\Phi'} = \sqrt{G_{00}'}\sqrt{G_{ij}'-\frac{G_{0i}'G_{0j}'}{G_{00}'}}e^{-2\Phi}G_{00} = \sqrt{G_{00}}e^{-2\Phi}\sqrt{G_{ij} - \frac{G_{0i}G_{0j}}{G_{00}}}.
\end{equation}
Thus we arrive at (also using $R \to \alpha'/R$)
\begin{align}
S \sim \int d^{D-1}x \sqrt{G_{ij} - \frac{G_{0i}G_{0j}}{G_{00}}}\sqrt{G_{00}}e^{-2\Phi} \nonumber \\
\left(G'^{ij}\partial_{i}T_{w}\partial_{j}T_{w}^{*}+\frac{w^2R^2G'^{00}}{\alpha'^2}T_{w}T_{w}^{*}+ G'^{0i}\frac{iwR}{\alpha'}\left(T_{w}\partial_{i}T_{w}^{*}- T_{w}^{*}\partial_{i}T_{w}\right)+ m^2T_{w}T_{w}^{*}\right).
\end{align}
Note that this field theory action indeed couples to the background NS-NS field. From here on we set $\Phi = constant$ (as we have done in all the other cases as well). \\
The term corresponding to $G'^{0i}$ in brackets needs to be written in the form $T^{*}\hat{\mathcal{O}}T$ to apply the manipulations as in section \ref{qm}. This gives schematically
\begin{equation}
\sqrt{G}G'^{0i}\left(T_{w}\partial_{i}T_{w}^{*}- T_{w}^{*}\partial_{i}T_{w}\right) \to -\partial_{i}\left(\sqrt{G}G'^{0i}\right)T_{w}^{*}T_{w} - 2\sqrt{G}G'^{0i}T_{w}^{*}\partial_{i}T_{w}.
\end{equation}
The first term is a spatial derivative term and is similar to $K(x)$ in section \ref{qm}. We thus include it in $K(x)$ and we ignore it in what follows. \\
Further following the logic from section \ref{qm}, we seek a path integral description of a system with Hamiltonian
\begin{equation}
H = p_i p_j g^{ij} + m_{local}^2 + K(x) + V^{i}p_{i}
\end{equation}
where $V^{i} = 2G'^{0i}wR/\alpha'$.
The Euclidean Lagrangian corresponding to this Hamiltonian equals\footnote{Expression (23.A.22) in \cite{Weinberg:1996kr} with $A_{ab} = 2g_{ab}$ and $B_{a} = V_{a}$.}
\begin{align}
L_E &= \frac{1}{4}\dot{x}^{i}\dot{x}^{j}g_{ij} + i\frac{V_{i}\dot{x}^{i}}{2} - \frac{V_{i}V^{i}}{4} + K(x)+ m_{local}^2\\
 &= \frac{1}{4}\dot{x}^{i}\dot{x}^{j}\bar{G}'_{ij} + iw\bar{G}'_{ij}G'^{0j}\frac{R}{\alpha'}\dot{x}^{i} - \bar{G}'_{ij}G'^{0i}G'^{0j}\frac{w^2R^2}{\alpha'^2} + \frac{w^2R^2G'^{00}}{\alpha'^2} + m^2 +K(x)
\end{align}
where we denoted $\bar{G}'_{ij}$ as the inverse to the purely spatial matrix $G'^{ij}$. This is \emph{not} the same as $G'_{ij}$ since the latter results from inverting the complete $G'^{\mu\nu}$ matrix and then looking at the spatial components.
Using matrix algebra, one can show the following identities
\begin{align}
\bar{G}'_{ij} &= G_{ij} - \frac{G_{0i}G_{0j}}{G_{00}}, \\
\bar{G}'_{ij}G'^{0i}G'^{0j} - G'^{00} &= - G_{00}, \\
\bar{G}'_{ij}G'^{0j} &= -B_{0i}.
\end{align}
The Euclidean Lagrangian reduces to (for $w = \pm1$)
\begin{equation}
L_E = \frac{1}{4}\dot{x}^{i}\dot{x}^{j}\left(G_{ij} - \frac{G_{0i}G_{0j}}{G_{00}}\right) \mp iB_{0i}\frac{R}{\alpha'}\dot{x}^{i} + G_{00}\frac{R^2}{\alpha'^2} + m^2 + K(x).
\end{equation}
Setting $R = \frac{\beta}{2\pi}$, we see that the $\dot{x}^{i}$ term reduces precisely to that given in section \ref{exten} and the other terms remain as before, in agreement with section \ref{nonstat}. There is however one extra correction term that we included in $K(x)$. This term vanishes for zero NS-NS background.

\end{document}